\documentclass[preprintnumbers, floatfix, preprintnumbers, letterpaper, superscriptaddress,nofootinbib,aps]{revtex4-2}
\pdfoutput=1
\usepackage{graphicx}
\usepackage{microtype}
\usepackage{amsmath}
\usepackage{amssymb}
\usepackage{subfigure}
\usepackage{hyperref}
\usepackage{url}
\usepackage{xcolor}
\usepackage{color}
\usepackage{mathrsfs}
\usepackage{calrsfs}
\usepackage{amsfonts}
\usepackage{latexsym}
\usepackage{ragged2e}
\usepackage{epstopdf}
\usepackage{textcomp}
\usepackage{phaistos}
\usepackage[utf8]{inputenc}
\usepackage{ulem}
\usepackage{float}
\usepackage{natbib}

\makeatletter
\renewcommand\@makefnmark{\hbox{\@textsuperscript{\normalfont\color{purple}\@thefnmark}}}
\renewcommand\@makefntext[1]{%
  \parindent 1em\noindent
            \hb@xt@1.8em{%
                \hss\@textsuperscript{\normalfont\@thefnmark}}#1}
\makeatother

\usepackage{caption}
\DeclareCaptionJustification{justified}{\leftskip=0pt \rightskip=0pt \parfillskip=0pt plus 1fil}
\definecolor{vividviolet}{rgb}{0.62, 0.0, 1.0}
\definecolor{amaranth}{rgb}{0.9, 0.17, 0.31}
\definecolor{palatinateblue}{rgb}{0.15, 0.23, 0.89}
\definecolor{brightpink}{rgb}{1.0, 0.0, 0.5}
\definecolor{cornflowerblue}{rgb}{0.39, 0.58, 0.93}
\definecolor{deepcarminepink}{rgb}{0.94, 0.19, 0.22}
\definecolor{radicalred}{rgb}{1.0, 0.21, 0.37}

\hypersetup{ linktoc=all,
    colorlinks, linkcolor={palatinateblue},
    citecolor={brightpink}, urlcolor={amaranth}
}

\graphicspath{{Images/}}

\def\sideremark#1{\ifvmode\leavevmode\fi\vadjust{\vbox to0pt{\vss
 \hbox to 0pt{\hskip\hsize\hskip1em
 \vbox{\hsize1.5cm\tiny\raggedright\pretolerance10000
 \noindent #1\hfill}\hss}\vbox to8pt{\vfil}\vss}}}%
\begin{document}

\title{Shadow of the Scalar Hairy Black Hole with Inverted Higgs Potential}

\author{Kok-Geng \surname{Lim}}
\email{K.G.Lim@soton.ac.uk}
\affiliation{Smart Manufacturing and Systems Research Group, University of Southampton Malaysia, 79100 Iskandar Puteri, Malaysia.}

\author{Xiao Yan \surname{Chew}}
\email{xiao.yan.chew@just.edu.cn (corresponding author)}
\affiliation{School of Science, Jiangsu University of Science and Technology, 212100, Zhenjiang, China.}

\begin{abstract}
We study the imaging of a hairy black hole (HBH) in the Einstein-Klein-Gordon theory, where Einstein gravity is minimally coupled to a scalar potential $V(\phi)=-\Lambda \phi^4 + \mu \phi^2$ with $\Lambda$ and $\mu$ are constants. As a consequence, a nontrivial scalar field at the event horizon $\phi_H$ allows the HBH to evade the no-hair theorem, bifurcate from the Schwarzschild black hole by acquiring some new properties, which can affect the shadow of the HBH received by a distant observer. The framework of ray-tracing is adopted to investigate the optical appearance of the HBH, thus the trajectories of light rays around the HBH can be classified into three emissions: direct, lensed and photon ring. Employing three models of optically and geometrically thin accretion disk, we compare the differences between the Schwarzschild black hole and HBH with same horizon radius in a specific model, and find that the size of the shadow and accretion disk increases as $\phi_H$ increases, but the brightness of the rings remain nearly unaffected, this implies our HBH can potentially mimic the Schwarzschild black hole if we vary the horizon radius of the HBH. Finally, we also constraint the parameter $\Lambda$ from the observations of supermassive black holes in the galactic center of M87 and Sgr A$^{*}$, which could offer new insights for imaging of black holes and astrophysical observations.
\end{abstract}

\maketitle

\section{Introduction}

According to the no-hair theorem, the state of a black hole in General Relativity (GR) can only be described by three physical quantities: mass, electric charge, and angular momentum \cite{Israel:1967wq,Carter:1971zc,Ruffini:1971bza}. However, the no-hair theorem can be circumvented, where a hairy black hole (HBH) can be typically bifurcated from the electrovacuum black holes such as the Schwarzschild black hole, Reissner-Nordstrom black hole, etc when a matter field is nontrivial at the event horizon of a black hole. Thus, an HBH can also be described by additional physical quantities which known as the  ``hair", and exhibit new phenomena in the strong gravity regime compared to the electrovacuum black holes, but they can become indistinguishable in the weak gravity regime.  

Recently, a class of scalar HBHs has been numerically constructed in Einstein-Klein-Gordon (EKG) theory, where the no-hair theorem can be circumvented by some non-positive definite scalar potentials $V(\phi)$ which minimally coupled with Einstein gravity. For instance, $V(\phi)$ with asymmetric vacua \cite{Corichi:2005pa, Chew:2022enh, Chew:2024rin} describes the first-order phase transition of our universe \cite{Coleman:1980aw}; $V(\phi)=-\Lambda \phi^4 + \mu \phi^2$ \cite{Gubser:2005ih, Chew:2023olq} and $V(\phi)=-\alpha^2 \phi^6$ \cite{Chew:2024evh}. The circumvention of no-hair theorem can be realized since $V(\phi)<0$ in some regions of $\phi$ can violate the weak energy condition, which is one of the conditions imposed by the no-hair theorem. For instance, the negative region of $V(\phi)$ with asymmetric vacua lies between two zeros of $V(\phi)$; $V(\phi)=-\Lambda \phi^4 + \mu \phi^2<0$ when $|\phi| > \sqrt{\mu/\Lambda} $; $V(\phi)=-\alpha^2 \phi^6<0$ for $-\infty<\phi<\infty$. Other constructions of HBHs in EKG can be found in Refs. \cite{Bechmann:1995sa,Dennhardt:1996cz,Bronnikov:2001ah,Martinez:2004nb,Nikonov:2008zz,Anabalon:2012ih,Stashko:2017rkg,Gao:2021ubl,Karakasis:2023ljt,Atmaja:2023ing,Li:2023tkw,Rao:2024fox,Wijayanto:2022bwx,Wijayanto:2023wru}. Besides, several classes of scalar HBHs has been constructed in the EKG theory with a real scalar field, which is minimally coupled with a U(1)-gauged complex scalar field associated with some potentials \cite{Kunz:2023qfg,Kunz:2024uux,Brihaye:2024mlm} and Proca field associated with a potential \cite{Herdeiro:2024pmv}.

Furthermore, some great achievements on the detections of astrophysical black holes have been witnessed in these recent years, in particular, the imaging of two supermassive black holes in the galactic center of Sgr A$^*$ and M87 by the Event Horizon Telescope (EHT) \cite{EventHorizonTelescope:2021dqv,EventHorizonTelescope:2022wkp,EventHorizonTelescope:2022xqj} not only confirm the predictions of black holes by GR in the strong gravity regime, but also provides us a strong motivation to investigate the shadow cast by the black holes, where a lot of new predictions from the theoretical calculations can be possibly tested in the near future \cite{Johannsen:2016uoh,Mizuno:2018lxz,Younsi:2021dxe,Chen:2022scf,Vagnozzi:2022moj,Genzel:2024vou,Ayzenberg:2023hfw,Buoninfante:2024oxl}. The realistic modeling for the shadow of black holes generally requires the sophisticated numerical simulation of General Relativistic Magnetic Hydrodynamics \cite{McKinney:2012vh,Moscibrodzka:2015pda,Chatterjee:2020eqc,Ressler:2020voz} to describe the distribution of electromagnetic radiation during the turbulent process of matter accreted into the black holes. Nevertheless, the ray-tracing method proposed by Gralla \textit{et al} \cite{Gralla:2019xty} can be adopted to simplify the study if we assume the accretion disk is optically and geometrically thin, hence the optical appearance of the black holes can be completely described by the trajectories of light rays. 

However, one of the predictions by GR, namely the Kerr hypothesis still can be challenged since some HBHs can exhibit an intriguing phenomenon, where their shadows can fit well with the current observations  \cite{Herdeiro:2021lwl,deSa:2024dhj}. Recently the geodesics of test particles around the HBH with $V(\phi)=-\Lambda \phi^4 + \mu \phi^2$ has been studied in our latest work \cite{Chew:2023olq}. Thus, it motivates us to investigate further the role of the scalar field at the event horizon on the optical appearance and report some interesting results in this paper. Other investigations of imaging of other HBHs have been studied extensively \cite{Vagnozzi:2019apd,Banerjee:2019nnj,Cunha:2019ikd,Konoplya:2019sns,Peng:2020wun, Zeng:2020vsj, Kumar:2020owy, Guo:2020zmf,Zeng:2020dco,Li:2021riw, Guerrero:2021ues,Khodadi:2021gbc, Yang:2022btw,Lee:2021sws,Li:2023djs,Li:2024kyv,Guerrero:2023sra,Wang:2024lte,Sui:2023tje,Xu:2024gjs,Huang:2024bbs,Hou:2024qqo,Meng:2023htc,Zeng:2023zlf,Tu:2024huh,Chen:2024aaa,Chen:2024cxi,Wei:2024cti,Myung:2024pob,KumarWalia:2022aop,Zheng:2024brm,Wang:2023rjl,Raza:2025pgs,Wu:2024wxo,Endo:2024opp,Pantig:2024ixc,Chen:2024luw,Meng:2025ivb,Gyulchev:2024iel,Macedo:2024qky,Zeng:2024ptv}.  

Besides, some horizonless compact objects can also exhibit this similar phenomenon, although they don't possess an event horizon \cite{Herdeiro:2021lwl,Bouhmadi-Lopez:2021zwt,Rosa:2022tfv,deSa:2024dhj,Rosa:2023qcv}. This indicates that the formation of the optical appearance of compact objects doesn't depend on the existence of an event horizon, but is mainly determined by the photon sphere of compact objects, where a portion of light rays can wind infinitely many times in a circular orbit in principle. Hence, it is actually a basic manifestation of the bending of light rays in a curved spacetime.

The content of our paper is organized as follows. In Sec.~\ref{sec:th}, we briefly present the numerical construction of our HBH model, including the EKG theory, the metric Ansatz, the derivation of the set of coupled differential equations from the equations of motion, the asymptotic behaviour of the functions and the properties of the HBH. In Sec~\ref{sec:geo}, we adopt the ray-tracing method to investigate the shadow of the HBH. In this formalism, we derive the effective potential of light rays from the radial geodesics equation. We characterize the light rays into three main emissions based on the number of intersections of the light rays with the accretion disk, and then adopt three models of the accretion disk to study the optical appearance of the HBH. In Sec~\ref{sec:res}, we present and discuss our numerical findings by comparing the optical appearance of the Schwarzschild black hole and the HBH. Finally, in Sec.~\ref{sec:con}, we summarize our work and present an outlook.

\section{Theoretical Setting} \label{sec:th}

\subsection{Theory and Ans\"atze}

In the Einstein-Klein-Gordon (EKG) theory, we consider the Einstein gravity minimally coupled with a scalar potential $V(\phi)$ of a scalar field $\phi$,
\begin{equation} \label{EHaction}
 S=  \int d^4 x \sqrt{-g}  \left[  \frac{R}{16 \pi G} - \frac{1}{2} \nabla_\mu \phi \nabla^\mu \phi - V(\phi) \right]  \,,
\end{equation}
where
\begin{equation} \label{vpot}
 V(\phi) = -\Lambda \phi^4 + \mu \phi^2 \,,
\end{equation}
with $\Lambda$, $\mu$ are the real-valued constants. Here $V(\phi)$ contains a false vacuum at $\phi=0$ and two degenerate maxima at $\phi_\text{max}=\pm \sqrt{\mu/(2\Lambda)}$, hence the profile of $V(\phi)$ is symmetry. It has been employed to construct the solutions of HBH which can emerge from the Schwarzschild black hole when $\phi$ is nontrivial at the horizon \cite{Chew:2023olq}. Moreover, the HBH can be smoothly connected with its counterpart gravitating scalaron in the small horizon limit \cite{Chew:2024bec}. Note that the gravitating scalaron can possess positive mass, this demonstrates that the inclusion of phantom field, which is associated with the Higgs-like potential $(\Lambda<0,\mu<0)$ for the construction of gravitating scalaron with negative mass can be avoided \cite{Dzhunushaliev:2008bq}. Recently Eq.~\eqref{vpot} has been employed to study the formation of crunch singularity behind the apparent horizon in the collapse of the Anti-de Sitter (AdS) bubble to demonstrate that the lower bound of $V(\phi)$ isn't a necessary condition for the occurrence of its formation \cite{Lozanov:2024eyb}. Nevertheless, $V(\phi)$ can be asymmetric by possessing two vacuums which are the false and true vacuums when a cubic term $\phi^3$ is inserted into $V(\phi)$. The corresponding asymmetric $V(\phi)$ has also been employed to construct the neutral HBH \cite{Chew:2022enh},  electrically charged HBH \cite{Chew:2024rin} and fermionic star\cite{DelGrosso:2023dmv,Berti:2024moe}. 

The Einstein and Klein-Gordon (KG) equations can be obtained, respectively by varying Eq.\eqref{EHaction}
\begin{equation} 
 R_{\mu \nu} - \frac{1}{2} g_{\mu \nu} R =  8\pi G  \left(  \nabla_\mu \phi \nabla_\nu \phi -    \frac{1}{2} g_{\mu \nu} \nabla_\alpha \phi \nabla^\alpha \phi - g_{\mu \nu}  V(\phi)  \right) \,,  \quad  \nabla_\mu \nabla^\mu \phi  =  \frac{d V}{d \phi} \,.  
\end{equation}
A static and spherically symmetric Ansatz can be employed to construct the solutions of HBHs,
\begin{equation}  \label{line_element}
ds^2 = - N(r) e^{-2 \sigma(r)} dt^2 + \frac{dr^2}{N(r)} + r^2  \left( d \theta^2+\sin^2 \theta d\varphi^2 \right) \,, 
\end{equation}
where $N(r)=1-2 m(r)/r$ with $m(r)$ is the Minser-Sharp mass function. Note that $m(\infty)=M$ which is the Arnowitt-Deser-Misner (ADM) mass. The substitution of Eq.\eqref{line_element} into the equations of motion yields a set of nonlinear ODEs as shown below,
\begin{equation}
m' = \frac{\beta}{4} r^2 \left( N \phi'^2 + 2 V \right) \,, \quad \sigma' = - \frac{\beta}{2} r \phi'^2 \,,    \quad
\left(  e^{- \sigma} r^2 N \phi' \right)' = e^{- \sigma} r^2  \frac{d V}{d \phi} \,,
\end{equation}
where the prime $(')$ denotes the derivative of the functions with respect to the radial coordinate $r$. The globally regular solutions of HBH can be constructed by directly solving the above ODEs, but it could be very challenging to obtain such closed form of HBH solutions when $V(\phi)$ is nontrivial, hence we integrate the above ODEs numerically from the horizon $r_H$ to infinity by using two ODE solver packages: Colsys \cite{Ascher:1979iha} and Matlab bvp4c \cite{kierzenka2001bvp}. Colsys adopts the Newton-Raphson method to tackle the boundary value problems for the nonlinear ODEs, incorporating adaptive mesh refinement to produce solutions with more than 1000 points to ensure high precision and provide estimation of error. Besides, bvp4c is a boundary value problem solver with adaptive meshing, based on the three-stage Lobatto IIa collocation method. 

In the numerics, we compactify the radial coordinate $r$ by $x=1-r_H/r$ with $x \in [0,1]$ where the numerical values $0$ and $1$ represent the horizon and infinity, respectively. Moreover, the boundary conditions of the functions at the horizon can be described in the form of power series expansion, which are given by a few leading terms:
\begin{align}
 m(r) &= \frac{r_H}{2}+ m_1 (r-r_H) + O\left( (r-r_H)^2 \right) \,, \\
\sigma(r) &= \sigma_H + \sigma_1   (r-r_H) + O\left( (r-r_H)^2 \right)  \,, \\
 \phi(r) &= \phi_H +  \phi'(r_H)  (r-r_H) + O\left( (r-r_H)^2 \right)  \,,
\end{align} 
where
\begin{equation}
   m_1 = \frac{\beta r^2_H}{2}  V(\phi_H)  \,, \quad  \sigma_1 = -  \frac{\beta r_H}{2} \phi'(r_H) \,, \quad   \phi'(r_H)= \frac{r_H \frac{d V(\phi_H)}{d \phi}}{1-\beta r_H^2 V(\phi_H)}  \,.
\end{equation} 
Here $\sigma_H$ and $\phi_H$ are the values of $\sigma(r)$ and $\phi(r)$ at the horizon, respectively. The denominator of $\phi'(r_H)$ should satisfy the condition $1-\beta r_H^2 V(\phi_H) \neq 0$ in order to keep the functions $\sigma(r)$ and $\phi(r)$ finite at the horizon. At the infinity, the functions have to satisfy the asymptotic flatness condition by demanding $m(\infty)=M$, $\sigma(\infty)=\phi(\infty)=0$. In addition, there are some free parameters in our numerics, which are $\sigma_H$, $\phi_H$, $M$, $\Lambda$, $r_H$, and $\mu$. In particular, the values $\sigma_H$, $\phi_H$ and $M$ can be obtained exactly when the boundary conditions are satisfied. Thus, we can introduce some dimensionless parameters in the numerics via $r \rightarrow r/\sqrt{\mu}$, $m \rightarrow  m/\sqrt{\mu}$, $\phi \rightarrow \phi/\sqrt{8 \pi G}$, $\Lambda \rightarrow 8 \pi G \Lambda \mu$. Hence, we are left with two parameters, which are $r_H$ and $\Lambda$ to describe the HBH.

\subsection{The Properties of HBH} \label{sec:prop}

\begin{figure}
\centering
\mbox{
(a)
\includegraphics[angle =0,scale=0.58]{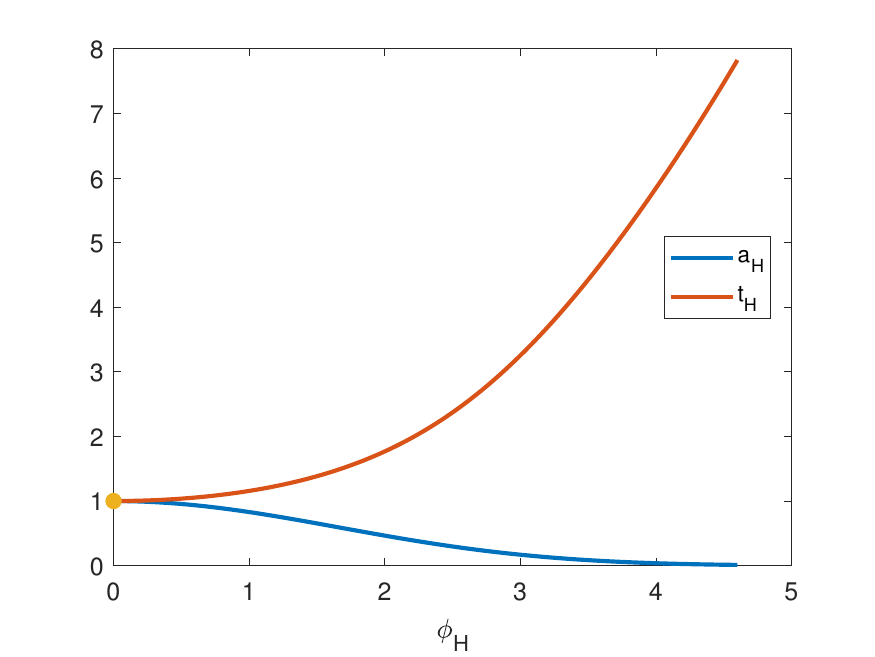}
(b)
\includegraphics[angle =0,scale=0.58]{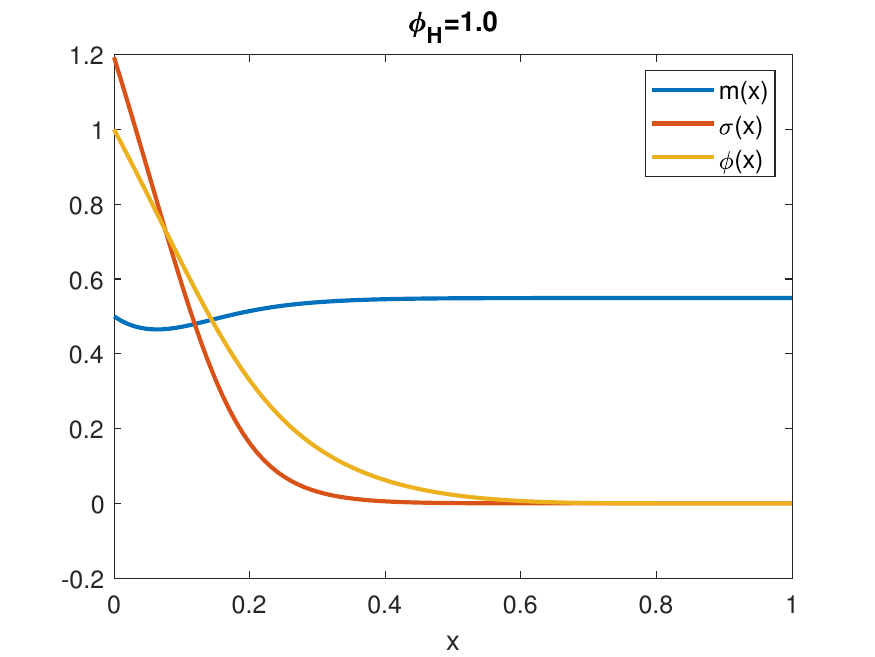}
}
\caption{(a) Two basic properties of the HBH: The reduced area of horizon $a_H$ and the reduced Hawking temperature $t_H$ as the function of $\phi_H$ for HBHs with $r_H=1$. The orange marker denotes the Schwarzschild black hole. (b) Typical profiles for the solution of the HBH with $r_H=1$, $\phi_H=1.0$ in the compactified coordinate $x$.}
\label{plot_basic_properties}
\end{figure}

The Schwarzschild black hole is the trivial solution to the ODEs when $\phi_H=0$ at the horizon. However, the HBH can emerge from the Schwarzschild black hole when $\phi_H>0$ at the horizon. In order to measure quantitatively how much the deviation of the HBH from the Schwarzschild black hole, we introduce two ``reduced" quantities at the horizon to describe the properties of the HBH: $a_H$ is the reduced area of horizon and $t_H$ is the reduced Hawking temperature,
\begin{equation}
 a_H = \frac{A_H}{16 \pi M^2} \,, \quad t_H = 8 \pi T_H M \,,
\end{equation}
where $A_H$ is the area of horizon of the HBHs, $T_H$ is the Hawking temperature, which are defined by
\begin{equation}
T_H = \frac{1}{4 \pi} N'(r_H) e^{-\sigma_H} \,, \quad A_H = 4 \pi r^2_H \,,
\end{equation}
To generate the solutions of HBH, we fix $r_H=1$ and then vary $\phi_H$. Recall that $a_H=t_H=1$ for the Schwarzschild black hole when $\phi_H=0$, as represented by an orange dot in Fig.~\ref{plot_basic_properties}. As $\phi_H$ increases from zero, $a_H$ decreases from 1 and eventually approaches zero. In contrast, $t_H$ monotonically increases from 1 \cite{Chew:2023olq}. 

Fig.~\ref{plot_basic_properties} (b) shows the typical profiles for the solutions of HBH in the compactified coordinate $x$. The functions $\sigma(x)$ and $\phi(x)$ decrease monotonically to zero but the function $m(x)$ decreases to a minimum value and then increases again to reach an almost constant value. The gradient of these profiles at the horizon is steeper than the gradient of the profiles of functions at the horizon for the HBH with asymmetric vacua \cite{Chew:2022enh,Chew:2024rin}, where they possess two different almost constant functions at the horizon and infinity which are connected by a sharp boundary.

In principle, the range of $\phi_H$ is $0 \leq \phi_H < \infty$. Besides, we find that the parameter $\Lambda$ is inversely proportional to $\phi_H$, as shown in the Table.~\ref{tab:my_label}. Thus, the separation between degenerate global maxima of $V(\phi)$ increases when $\phi_H$ increases, since $|\phi_\text{max}|=\sqrt{\mu/(2\lambda)}$. 

\begin{table}[H]
    \centering
    \begin{tabular}{|c|c|c|c|c|}
        \hline
        $\phi_H$ & 0.5 & 1.0 & 2.0 & 3.0 \\
        \hline
        $\Lambda$ & 13.4344 & 3.4728 & 1.0012 & 0.5713 \\
        \hline
    \end{tabular}
    \caption{Some values of $\Lambda$ correspond to $\phi_H$ with $r_H=1$.}
    \label{tab:my_label}
\end{table}

\section{The Ray-Tracing Method} \label{sec:geo}

In the framework of ray-tracing to study the shadow cast by the HBH \cite{Gralla:2019xty, Guerrero:2021ues, Guerrero:2023sra}, we consider some models of accretion disk which are optically and geometrically thin, surround and lie on the equatorial plane of the HBH \cite{Gralla:2019xty}. We assume that the accretion disk itself does not absorb light rays. Instead, the brightness of light rings can be enhanced each time when light rays cross the accretion disk. The geometrically thin nature of the accretion disk assumes that its width is far smaller than its size of radius, this implies that the light rays can propagate through the accretion disk without being blocked. Moreover, an observer at a far distance from the HBH could receive the image of the HBH which appears face-on, since the disk lies on the equatorial plane, which is perpendicular to the observer.

Therefore, the trajectories of light rays play a crucial role for the optical appearance of the HBH, which can be fully described by the geodesics equation. Hence, we begin our study with the Lagrangian which is given by
\begin{equation}
\label{eq_lagrangian}
  \mathcal{L} =  \frac{1}{2} g_{\mu \nu} \dot{x}^\mu \dot{x}^\nu = \frac{1}{2} \left( -N e^{-2 \sigma} \dot{t}^2 + \frac{\dot{r}^2}{N} + r^2 \dot{\theta} + r^2 \sin{\theta}^2 \dot{\varphi}^2 \right) = \epsilon \,,
\end{equation} 
where $\epsilon=0$ refers to massless particles and $\epsilon=-1$ refers to massive particles. The dot denotes the derivative of a function with respect to the affine parameter. 

Since the spacetime of HBH is static and stationary, which gives rise to the existence of two conserved quantities: energy $E$ and angular momentum $L$,
\begin{equation}
\label{eq_conserved}
 E = - \frac{\partial \mathcal{L}}{\partial \dot{t}} = e^{-2 \sigma} N \dot{t} \,, \quad  L =  \frac{\partial \mathcal{L}}{\partial \dot{\varphi}} = r^2 \dot{\varphi} \,. 
\end{equation}
The spherically symmetric property of the spacetime allows us to concentrate on the motion of test particles on the equatorial plane $(\theta=\pi/2)$, hence this yields the radial equation $\dot{r}$,
\begin{equation} \label{rad_eqn}
 e^{-2 \sigma} \dot{r}^2 =    E^2 -  V_{\text{eff}}(r)  \,, 
\end{equation}
where the effective potential $V_{\text{eff}}(r)$ is given by
\begin{equation}
 V_{\text{eff}}(r) = e^{-2 \sigma} N \left(  \frac{L^2}{r^2} - \epsilon \right) \,.
\end{equation}
If $V_{\text{eff}}(r)$ contains an inflection point, this implies that the test particle is located in the innermost stable circular orbit (ISCO) with the radius $r_\text{ISCO}$, thus it can't be captured by the HBH and escape to the infinity, where it can be determined by the condition $V'_{\text{eff}}(r)=V''_{\text{eff}}(r)=0$, and their explicit forms are given by
\begin{align}
  V'_{\text{eff}}(r) &=  \left( \frac{N'}{N} - 2 \sigma'  \right)  V_{\text{eff}} - \frac{2 L^2 e^{-2 \sigma } N}{r^3}    \,, \label{ISCOe1} \\
 V''_{\text{eff}}(r) &= \left(  \frac{N''}{N} - \frac{4 \sigma' N'}{N} + 4 N \sigma'^2 - 2 \sigma''  \right)    V_{\text{eff}} + 2 L^2 e^{-2\sigma} \left( \frac{3 N}{r^4} + \frac{4 N \sigma'}{r^3} - \frac{2 N'}{r^3}  \right) \,. \label{ISCOe2}
\end{align}
In particular for the case of light rays $(\epsilon=0)$, we can define $b \equiv L/E$, namely, the impact parameter in the radial equation $\dot{r}$,
\begin{equation}
\label{eq_erdot}
 e^{-2 \sigma} \dot{r}^2 =  E^2 \left( 1 - b^2  v_{\text{eff}}(r) \right) \,,
\end{equation}
where 
\begin{equation}
 v_{\text{eff}}(r) =  \frac{e^{-2 \sigma} N}{r^2} \,.
\end{equation}
A portion of light rays can travel in a circular orbit in the vicinity of the HBH, namely, the photon sphere, where $v_{\text{eff}}(r)$ possesses an extremum at a point $r_p$, which can be determined by the condition $v'_{\text{eff}}(r_p)=0$. The photon sphere is stable if $v_{\text{eff}}(r)$ possesses a local minimum but unstable if $v_{\text{eff}}(r)$ possesses a local maximum. Since $\dot{r}=0$ for photon sphere \cite{Chandrasekhar:1985kt, berti2014black}, allows us to define the critical impact parameter $b_c$, which corresponds to the photon sphere, as 
\begin{equation}
    b_c = \frac{r_p e^{\sigma(r_p)}}{\sqrt{N\left(r_p\right)}} \,.
\end{equation}

The trajectories of light rays around the HBH can be calculated by integrating the below expression \cite{Guerrero:2021ues, Rosa:2022tfv},
\begin{equation}
\label{eq_dphidr}
\frac{ \dot{\varphi} }{ \dot{r} } \equiv \frac{d \varphi}{dr} = \pm \frac{b}{r^2} \frac{1}{e^\sigma \sqrt{ 1 - b^2 v_{\text{eff}}(r) }} \,, 
\end{equation}
where $\pm$ represents the outgoing and ingoing directions of the light rays, respectively. The range of integration is determined by the values of $b$. If $b<b_c$, then the total change of $\varphi$ is obtained by performing the below integration \cite{Gralla:2019xty, Peng:2020wun}:
\begin{equation}
\label{eq_dphi}
 \varphi = \int_{r_H}^{\infty} \frac{b}{r^2} \frac{dr}{e^\sigma \sqrt{ 1 - b^2 v_{\text{eff}}(r) }}  \,.
\end{equation}
Whereas for $b>b_c$, the total change of $\varphi$ is:
\begin{equation}
\label{eq_dphi2}
 \varphi = 2 \int_{r_\text{min}}^{\infty} \frac{b}{r^2} \frac{dr}{e^\sigma \sqrt{ 1 - b^2 v_{\text{eff}}(r) }}  \,,
\end{equation}
where $r_\text{min}$ is the minimal radial distance from a light ray to the HBH, as long as $1 - b^2 v_{\text{eff}}(r) > 0$. Then the trajectories of light rays can be clearly visualized by presenting them in the two-dimensional Cartesian coordinate $(x,y)$ by
\begin{equation}
    x = r \cos(\varphi) \,, \quad y=r \sin(\varphi)\,.
\end{equation}

\subsection{The Characterization of Light Rays}

The total number of orbits, $n=\varphi/(2 \pi)$ for the light rays revolving around the HBH is defined as $m$-intersections of light rays with the accretion disk on the equatorial plane \cite{Gralla:2019xty}, which allows us to classify the trajectories of light rays into three main categories as shown in the below: 
\begin{align}
\label{eq_categories}
    &\mbox{Direct emission} \, (m=1): 1/4 \leq n < 3/4 \quad \mbox{for} \quad b \in \left(0, b_2^{-} \right) \cup \left(b_2^{+}, \infty \right) \,, \\ \nonumber
    &\mbox{Lensed emission} \, (m=2): 3/4 < n < 5/4 \quad \mbox{for} \quad b \in \left(b_2^{-}, b_3^{-} \right) \cup \left(b_3^{+}, b_2^{+} \right) \,, \\ \nonumber
    &\mbox{Photon ring emission} \, (m=3): n > 5/4 \quad \mbox{for} \quad b \in \left(b_3^{-}, b_3^{+} \right) \,.
\end{align}

\begin{figure}
\centering
\mbox{
(a)
\includegraphics[angle =0, trim=20mm 0mm 5mm 0mm, clip, scale=0.6]{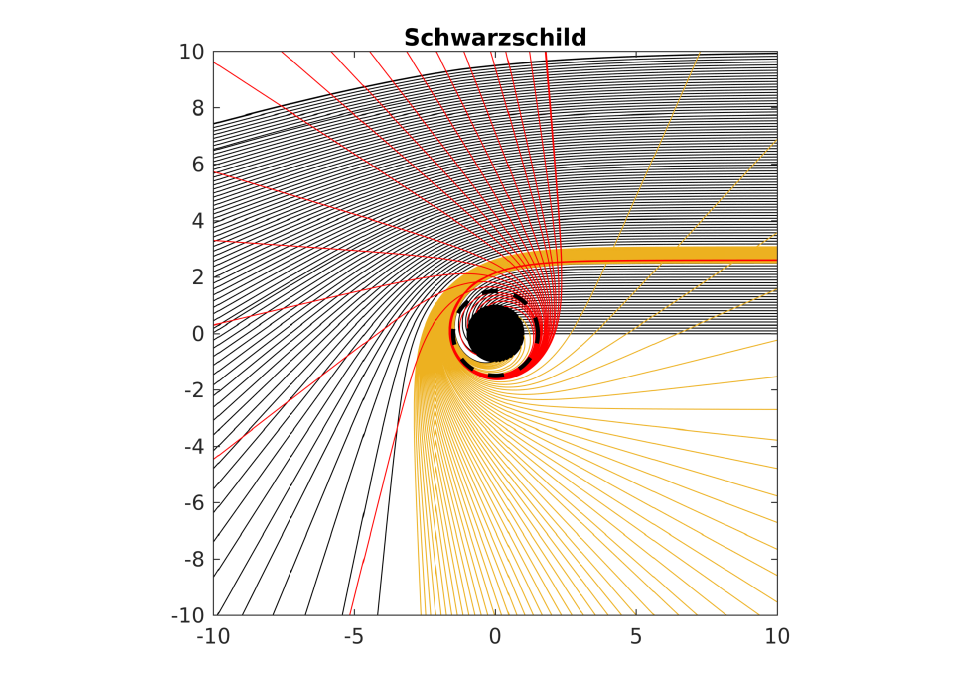}
(b)
\includegraphics[angle =0,scale=0.58]{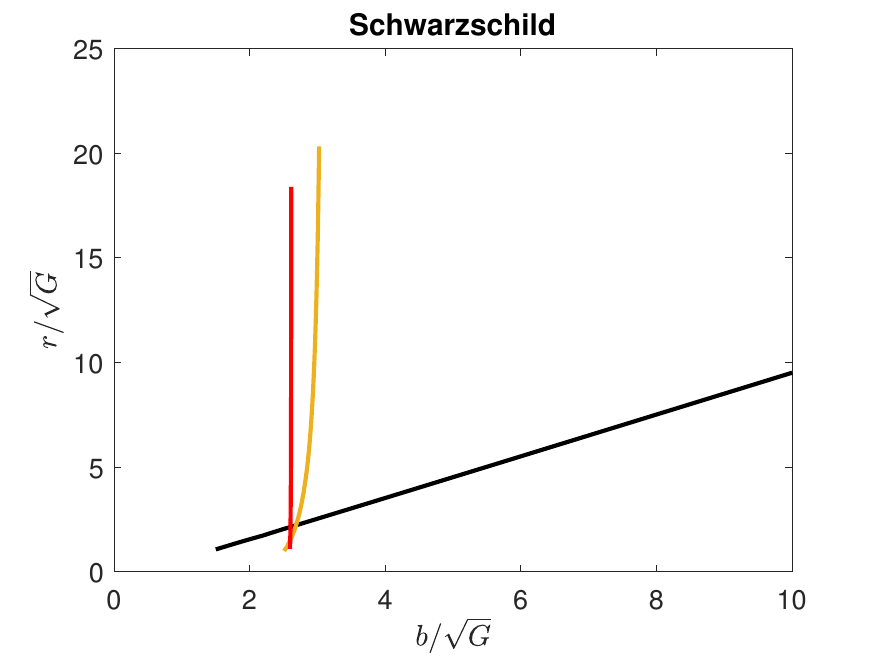}
}
\caption{
(a) The trajectories of light rays are presented in the two-dimensional Cartesian coordinate $(x,y)$ for the Schwarzschild black hole with $r_H=1$. A light ray is fixed by a value of impact parameter $b$, then the step size between the light rays for the direct (black), lensed (gold), and photon ring (red) emissions are fixed by 0.1, 0.01, and 0.001, respectively. The black hole and photon sphere are represented by the black disk and dashed circle, respectively. The accretion disk is represented by the vertical straight line $(0,y)$. (b) The three transfer functions: $r_1(b)$ (black), $r_2(b)$ (gold), and $r_3(b)$ (red) as a function of scaled impact parameter $b\sqrt{G}$ for the Schwarzschild black hole with $r_H=1$.
}
\label{plot_schwarzschild1}
\end{figure}

Note that $b_m^{-} < b_c$ and $b_m^{+} > b_c$. Thus, for instance, we can directly visualize the characterization of light rays into direct (black), lensed (gold) and photon ring (red) emissions around the Schwarzschild black hole with $r_H=1$ in Fig.~\ref{plot_schwarzschild1} (a) by numerically integrating Eq.~\eqref{eq_dphidr}. Suppose an observer is assumed to be located at the right border of the figure (north pole direction), which is far away from the black hole (black disk), and the accretion disk lies in a vertical straight line $(0,y)$. The light rays originated from the north pole direction and are specified by a specific value of $b$, hence they fall into the black holes when $b<b_c$, forming a dark region in the optical appearance known as the shadow, but scattered away from the black hole to be received by the observer when $b>b_c$ \cite{Gralla:2019xty}.

The direct emission of photon trajectories describes the light paths that cross the accretion disk only once, but dominate the optical appearance of the object by primarily contributing to the brightness and the width of the radiation ring, thus can reflect almost full features of the accretion disk. The lensed emission of photon trajectories describes the light paths that can cross the equatorial plane twice to provide a secondary contribution to the light. The photon ring emission trajectories describe that light can cross the accretion disk at least three times. However, the photon trajectories with higher order ($m > 3$) have a very small influence on the overall brightness, but are more sensitive to the background geometry. These higher-order paths are generally included within the photon ring due to their narrow range of impact parameters \cite{Gralla:2019xty, Peng:2020wun, Guerrero:2023sra}.

Based on the three categories of light rays, the solutions of orbital equation Eq.~\eqref{eq_dphidr} can be defined as the transfer function $r_m(\varphi,b)$ for the light rays \cite{Gralla:2019xty, Peng:2020wun}: 
\begin{align}
    r_1(b) &= r \left(\frac{\pi}{2}, b \right) \,, \ b \in (b_1^{-}, \infty)\,, \\ \nonumber
    r_2(b) &= r \left(\frac{3 \pi}{2}, b \right)\,, \ b \in (b_2^{-}, b_2^{+})\,, \\ \nonumber
    r_3(b) &= r \left(\frac{5 \pi}{2}, b \right)\,, \ b \in (b_3^{-}, b_3^{+}) \,.
\end{align}
As discussed in \cite{Gralla:2019xty}, the slope of $r_m(\varphi,b)$, which is given by $\frac{dr_m}{db}$, describes the demagnification factor on the intensity of the optical image for the HBH. Thus, a steeper slope of $r_m(b)$ implies a lesser contribution of light rays to the optical image of the HBH. As depicted in Fig.~\ref{plot_schwarzschild1} (b), $\frac{dr_m}{db}$ increases with $m$ for the Schwarzschild black hole with $r_H=1$. Hence, $r_1(b)$ produces the ``direct image'' of the disk. However, $r_2(b)$ generates a significantly demagnified image of the disk's backside, which is termed the ``lensing ring''. $r_3(b)$ yields an extremely demagnified image of the disk's front side, referred to as the ``photon ring''.  $r_m(b)$ with higher-order produces images which are so highly demagnified, hence they can be disregarded. Besides, the values of $b^\pm_m$ for the Schwarzschild black hole with $r_H=1$ can be found in the first row of Table~\ref{table_b}.

\subsection{Models of Accretion Disk}

\begin{figure}
\centering
\mbox{
(a)
\includegraphics[angle =0,scale=0.58]{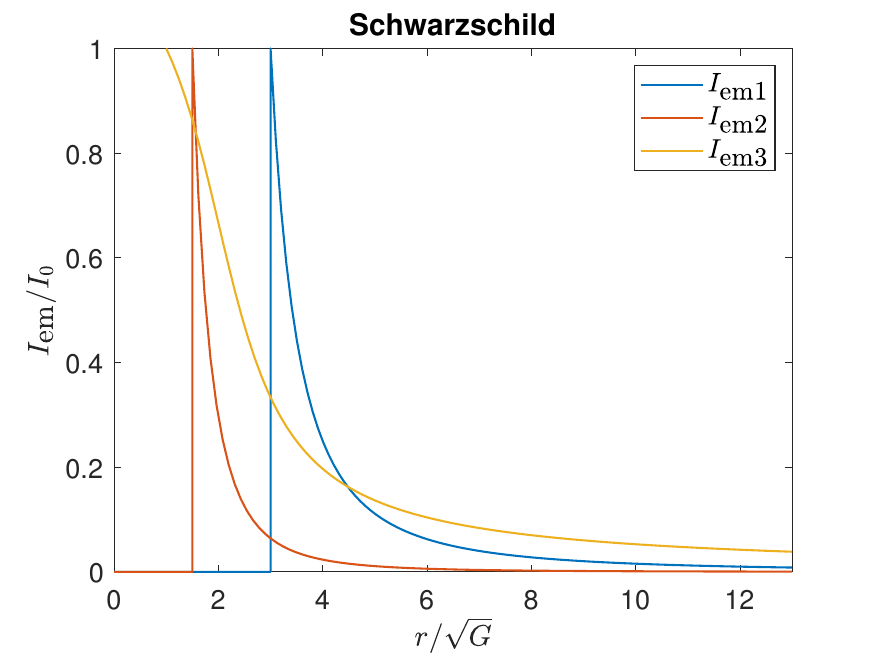}
(b)
\includegraphics[angle =0,scale=0.58]{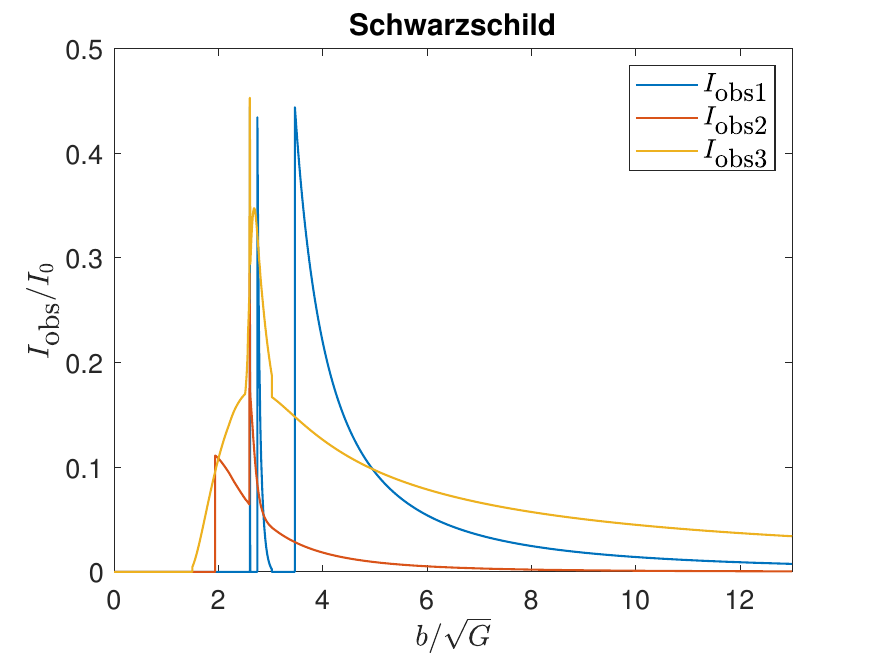}
}
\caption{(a) The three models of emitted intensity profiles: $I_\text{em1}(r)/I_0$, $I_\text{em2}(r)/I_0$, $I_\text{em3}(r)/I_0$ as a function of scaled $r\sqrt{\mu}$ for the Schwarzschild black hole with $r_H=1$. (b) The three models of observed intensity profiles: $I_\text{obs1}(r)/I_0$, $I_\text{obs2}(r)/I_0$, $I_\text{obs3}(r)/I_0$ as a function of scaled $b\sqrt{G}$ for the Schwarzschild black hole with $r_H=1$.
}
\label{Schwarzschild_Iobs_Iem}
\end{figure}

In order to model the formation of optical appearance of the black holes in the ray-tracing method, we consider that the shadow of the black holes can be cast by three specific models of emission intensity of the light rays from the accretion disk, which are given as \cite{Gralla:2019xty, Peng:2020wun, Zeng:2020vsj, Li:2021riw, Guerrero:2021ues, Rosa:2022tfv, Yang:2022btw, Guerrero:2023sra}:
\begin{align}
\mbox{Model I}: \quad 
    I_\text{em1}(r) &= 
        \begin{cases}
            I_0 \dfrac{1}{\left( r - (r_\text{ISCO} - 1) \right)^2}\,, &  \quad r> r_\text{ISCO}\\
            0\,, & \quad r \leq r_\text{ISCO} 
        \end{cases}
        \,,  \label{Model_I} \\
   \mbox{Model II}: \quad I_\text{em2}(r) &= 
        \begin{cases}
            I_0 \dfrac{1}{\left( r - (r_{p} - 1) \right)^3}\,, &  \quad r> r_{p}\\
            0\,, & \quad r \leq r_{p} 
        \end{cases}
        \,,  \label{Model_II} \\
   \mbox{Model III}: \quad I_\text{em3}(r) &= 
        \begin{cases}
            I_0  \dfrac{\frac{\pi}{2} - \arctan( r - (r_\text{ISCO} - 1 ))}{\frac{\pi}{2} - \arctan( r_H - (r_\text{ISCO} - 1 ))}\,, &  \quad r> r_{H}\\
            0\,, & \quad r \leq r_{H} 
        \end{cases}
        \,. \label{Model_III}
\end{align}
As shown in Fig.~\ref{Schwarzschild_Iobs_Iem} (a), the sharp spikes of $I_\text{em1}(r)$, $I_\text{em2}(r)$ and $I_\text{em3}(r)$ indicate that the emissions from the accretion disk around the Schwarzschild black hole with $r_H=1$ are located at $r_\text{ISCO}$, $r_p$ and $r_H$, respectively. This implies that the emission of $I_\text{em1}(r)$ only occurs when $r \leq r_\text{ISCO}$ and then gradually decreases to zero at infinity. The emission of $I_\text{em2}(r)$ behaves similarly with $I_\text{em1}(r)$ but only happens when $r \leq r_{p}$. The emission of $I_\text{em2}(r)$ only begins when $r>r_H$ but decays more smoothly to zero compared to the previous two models.

Due to the effect of gravitational redshift, the frequency of the light rays emitted by the source $\nu_\text{em}$ \cite{Gralla:2019xty} won't remain the same as the frequency received by the observer $\nu_\text{obs}$, but these frequencies can be related by a simple relation as shown in the below:
\begin{equation}
    \nu_\text{obs} = g^{1/2}_{tt} \nu_\text{em}.
\label{nu_obs}
\end{equation}
As the ratio, $I/\nu^3$ is conserved along the photon trajectory:
\begin{equation}
    \frac{I_{\text{obs},\nu_\text{obs}}}{\nu_\text{obs}^3} = \frac{I_{\text{em},\nu_\text{em}}}{\nu_\text{em}^3}.
\end{equation}
Hence, the observed intensity profiles are related to the emitted intensity profiles as follows:
\begin{equation}
    I_\text{obs} = \int I_{\text{obs}, \nu_\text{obs}} d \nu_\text{obs} = \int \frac{\nu_\text{obs}^3}{\nu_\text{em}^3} I_{\text{em},\nu_\text{em}} d \nu_\text{obs} = \int g_{tt}^{2} I_{em,\nu_\text{em}} d \nu_\text{em} = g_{tt}^2 I_\text{em},
\end{equation}
Since additional brightness will be acquired each time the light rays intersect with the accretion disk \cite{Gralla:2019xty}, the total observed intensity $I_\text{obs}(b)$ can be written as:
\begin{equation}
    I_\text{obs}(b) = \sum_m \left( N e^{-2 \sigma} \right)^2 I_\text{em} \Big|_{r=r_m(b)}\,. \label{Itotal}
\end{equation}

Hence, the effect of $I_\text{obs}(b)$ is demonstrated for the Schwarzschild black hole with $r_H=1$ in Fig.~\ref{Schwarzschild_Iobs_Iem} (b). For Model I, $I_\text{obs1}(r)/I_0$ is dominated by the direct emission, which corresponds to the outermost spike of $I_\text{obs1}(r)/I_0$ at $r=r_\text{ISCO}$. Similar to $I_\text{obs1}(r)/I_0$, $I_\text{obs1}(r)/I_0$ decreases from its maximum value to zero as $r \rightarrow \infty$. However, we could clearly observe the existence of two spikes in $I_\text{obs1}(r)/I_0$ when $r<r_\text{ISCO}$, due to the contributions from the lensed and photon ring emissions are included, where $I_\text{em1}(r)/I_0$ assumes no emission of light rays when $r<r_\text{ISCO}$. For Model II, $I_\text{obs2}(r)/I_0$ in Fig.~\ref{Schwarzschild_Iobs_Iem} (b) demonstrates an overlapping of three emissions for the Schwarzschild black hole with $r_H=1$ across a broader range of scaled $b \sqrt{G}$. Analogous to $I_\text{em2}(r)/I_0$, there is no emission for $I_\text{obs2}(r)/I_0$ when $r<r_p$, hence the innermost spike of $I_\text{obs2}(r)/I_0$ marks the initial emission, that corresponds to the direct emission, which decreases monotonically to zero when $r>r_p$. The sharper spike of $I_\text{obs2}(r)/I_0$ at the outermost actually represents a near-overlapping of lensing and photon ring emissions (zooming reveals a split) at an almost identical $b$ which is close to $b_c$. However, both emissions quickly diminish, leaving direct emission dominant again as $b$ increases. $I_\text{obs3}(r)/I_0$ of Model III is similar to Model II where the three emissions also highly overlap around $b=b_c$ but across a narrow range of scaled $b \sqrt{G}$.

\section{Results and Discussions}\label{sec:res}

\begin{figure}
\centering
\mbox{
(a)
\includegraphics[angle =0,scale=0.58]{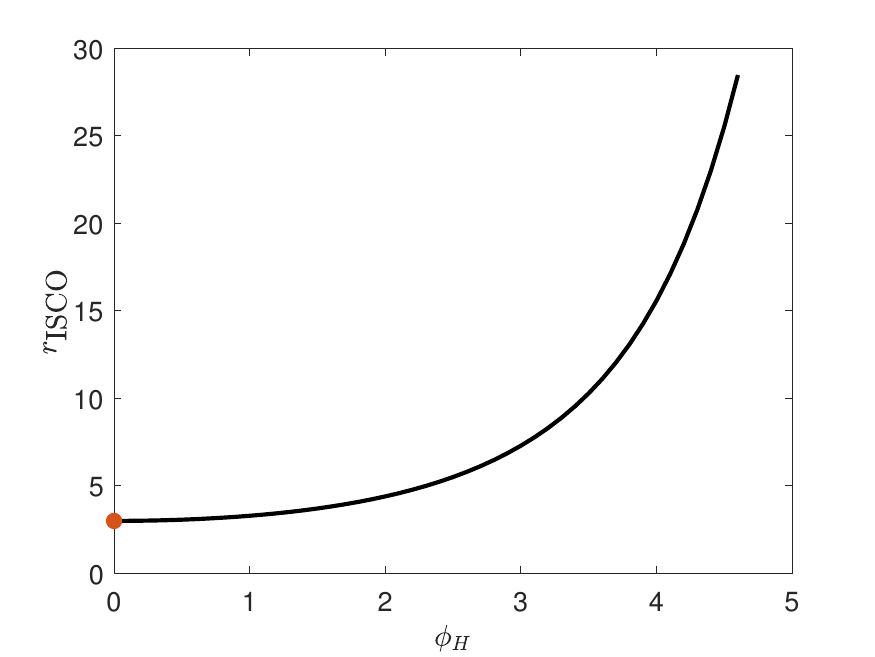}
(b)
\includegraphics[angle =0,scale=0.58]{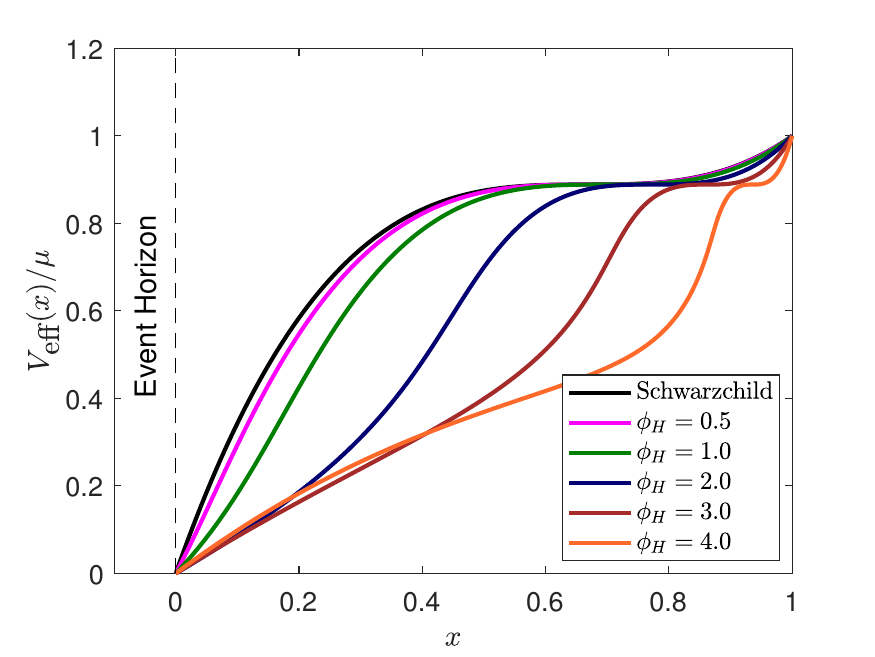}
}
\mbox{
(c)
\includegraphics[angle =0,scale=0.58]{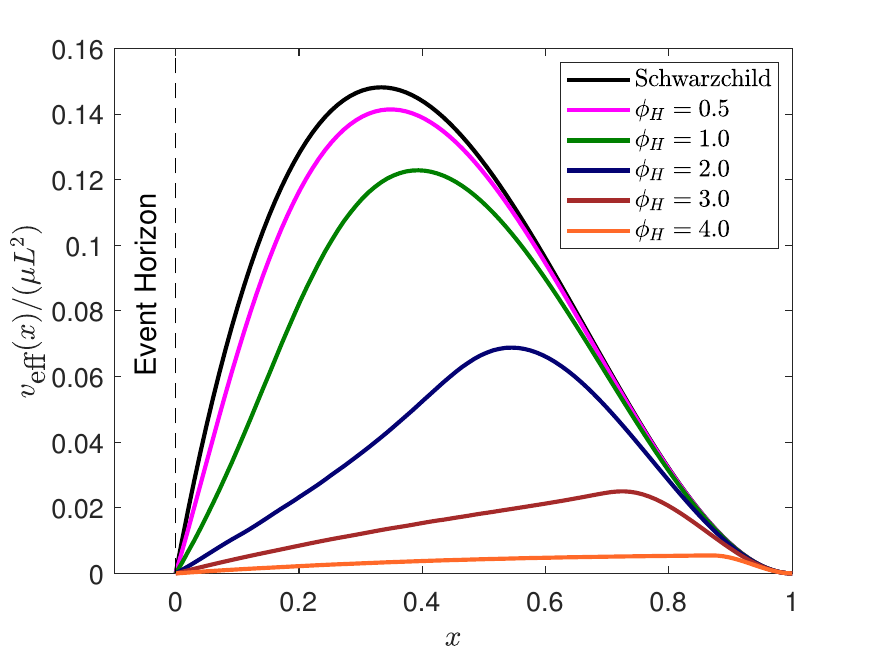}
(d)
\includegraphics[angle =0,scale=0.58]{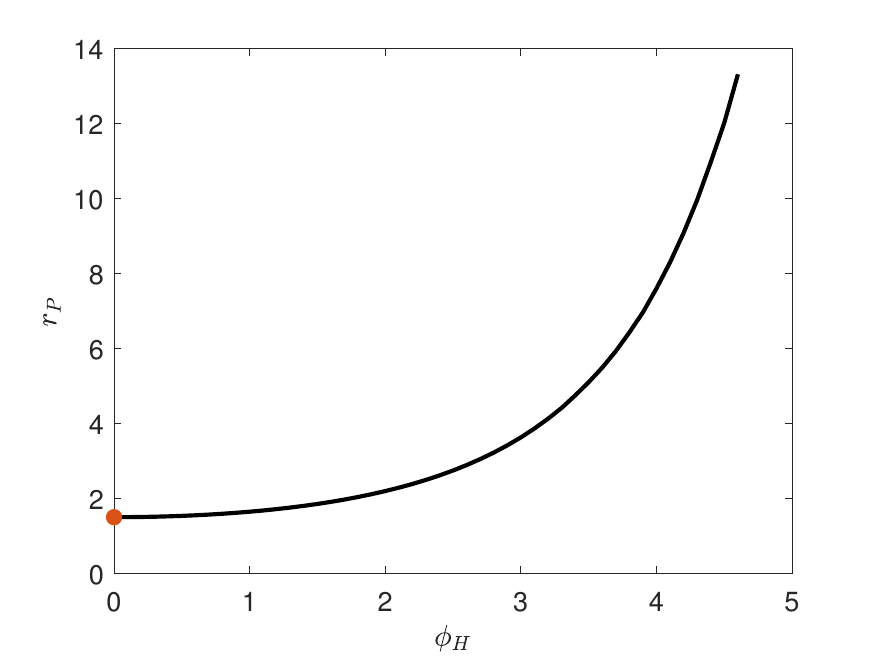}
}
\caption{(a) The location of ISCO, $r_\text{ISCO}$ as a function of $\phi_H$ for a massive test particle around the HBH with $r_H=1$. (b) The scaled effective potential $V_\text{eff}(x)/\mu$ of ISCO with several values of $\phi_H$ in the compactified coordinate $x$. (c) The scaled effective potential $v_\text{eff}(x)$ of the light rays in the compactified coordinate $x$ around HBHs with $r_H = 1$ and different values of $\phi_H$. (d) The location of the photon sphere, $r_p$ as a function of $\phi_H$. Note that the orange dot marks the reference point for the Schwarzschild black hole.}
\label{veff}
\end{figure}

We begin our investigation on the shadow of this HBH by analyzing $V_\text{eff}(r)$ of the test particles, in particular for the massive test particle, since the accretion disc is assumed to be located at the ISCO. Hence, the numerical solving of Eqs.~\eqref{ISCOe1} and \eqref{ISCOe2} could allow us to obtain $r_\text{ISCO}$ of a massive test particle, as shown in Fig.~\ref{veff} (a), where $r_\text{ISCO}$ bifurcated from the Schwarzschild black hole with $r_{\text{ISCO}}=3$  (orange dot) when $\phi_H>0$, and then increases monotonically when $\phi_H$ increases. Fig.~\ref{veff} (b) shows the scaled effective potential $V_\text{eff}(x)/\mu$ in the compactified coordinate $x$ for the test particle in the ISCO around the HBH with $r_H=1$ and different values of $\phi_H$, which contains the inflection point that corresponds to $V'_{\text{eff}}(r)=V''_{\text{eff}}(r)=0$, where it neither falls into the HBH nor flees to infinity.

\begin{figure}
\centering
\mbox{
\includegraphics[angle =0,scale=0.58]{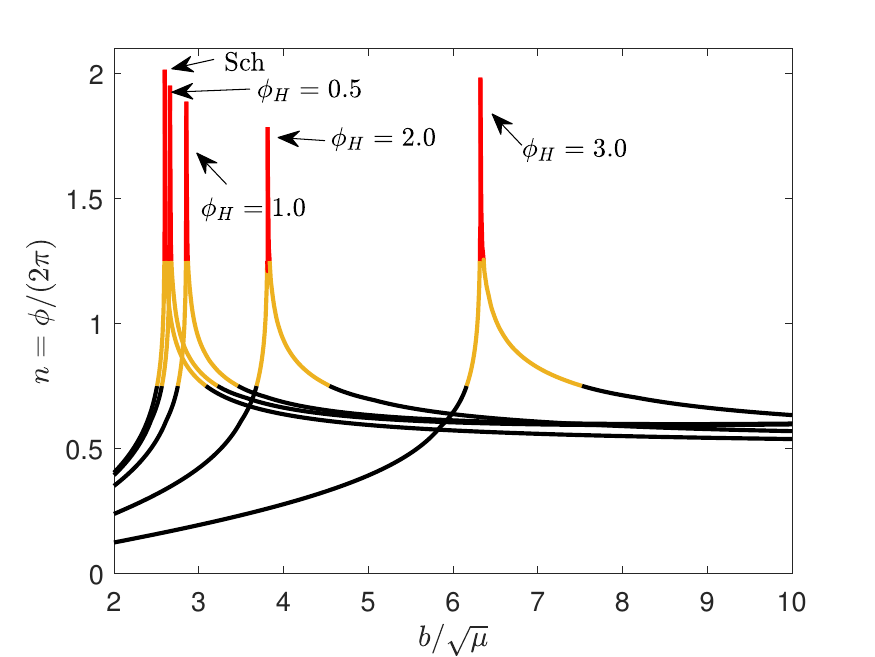}
}
\caption{The fractional number of orbits, $n=\varphi/(2 \pi)$ as a function of scaled impact parameter $b/\sqrt{\mu}$ for the Schwarzschild black hole with $r_H=1$ (labelled as Sch) and HBHs with $r_H=1$ and few values of $\phi_H$. The black, gold, and red curves represent three emissions of light rays: direct, lensed, and photon ring, respectively.}
\label{plot_orbits}
\end{figure}

Fig.~\ref{veff} (c) presents the scaled effective potential $v_\text{eff}(x)$ of light rays around HBHs with $r_H = 1$ for different values of $\phi_H$ in the compactified coordinate $x$. The presence of a peak in $v_\text{eff}(x)$ indicates the existence of an unstable photon sphere. As shown in Fig.~\ref{veff} (d), we observe that the height of the peak of $v_\text{eff}(x)$ decreases but its location $r_p$ deviates from the Schwarzschild black hole with $r_{\text{p}}=3/2$ (orange dot) and exhibits a monotonic increase with increasing values of $\phi_H$. 

Since $r_\text{ISCO}$ and $r_p$ of HBHs increase monotonically with the increment of $\phi_H$, then we can pick several values of $\phi_H=0.5, 1.0, 2.0, 3.0$ of the HBHs with $r_H=1$ to compare their optical appearance with the Schwarzschild black hole with $r_H=1$. In Fig.~\ref{plot_orbits}, analogous to the Schwarzschild black hole, the number of revolving of light rays $n$ around the HBHs with $r_H=1$ can still be described $n=\varphi/(2 \pi)$ as direct (black curve) with $1/4 \leq n <3/4$ for $b \in \left(0, b_2^{-} \right) \cup \left(b_2^{+}, \infty \right)$, lensed (gold curve) with $3/4<n<5/4$ for $b \in \left(b_2^{-}, b_3^{-} \right) \cup \left(b_3^{+}, b_2^{+} \right)$ and photon sphere (red curve) with $n>5/4$ for $b \in \left(b_3^{-}, b_3^{+} \right)$. However, when $\phi_H$ increases, the profile $n=\varphi/(2 \pi)$ for the light rays around the HBHs can deviate significantly from the Schwarzschild black hole with $r_H=1$, since the values of $b^\pm_m$ correspond to these three trajectories around the HBHs increase significantly from the Schwarzschild black hole, as demonstrated in Table~\ref{table_b}. Hence, the ranges of $b^\pm_m$ for these three types of trajectories also increase. Similarly, the location of the peak of $n$ is located approximately at $b_c$, which also increases when $\phi_H$ increases as shown in Table.~\ref{table_b}.

\begin{table}[H]
\begin{center}
\begin{tabular}{ |c|c|c|c|c|c|c| } 
 \hline
 $\phi_H$ &  $b_c$  & $b^{-}_1$ & $b^{-}_2$  & $b^{+}_2$  & $b^{-}_3$ & $b^{+}_3$ \\ 
\hline
 0 (Schwarzschild) & 2.59808 & 1.42385  & 2.50757  & 3.08378  & 2.5939  & 2.61396  \\
 \hline
 0.5 & 2.65914 & 1.44367   & 2.56385  & 3.22128  & 2.65479  & 2.67592  \\
\hline
 1.0 & 2.85289 & 1.55489  & 2.75255  & 3.46233  & 2.84853  & 2.87104  \\
 \hline
 2.0 & 3.81165 & 2.08403  & 3.68146 & 4.54288  & 3.80683  & 3.83529  \\
 \hline
 3.0 & 6.32173 & 3.70562   & 6.15937  & 7.52168  & 6.31513  & 6.35965  \\
 \hline
\end{tabular}
\caption{Some values of impact parameters $b^\pm_m$ correspond to three types of light rays: direct emission with $b \in \left(0, b_2^{-} \right) \cup \left(b_2^{+}, \infty \right)$; lensed emission with $b \in \left(b_2^{-}, b_3^{-} \right) \cup \left(b_3^{+}, b_2^{+} \right)$; and photon ring emission with $b \in \left(b_3^{-}, b_3^{+} \right)$ around the Schwarzschild black hole with $r_H=1$ and HBHs with $r_H=1$ and several $\phi_H$. Besides, $b_c$ is the critical impact parameter which corresponds to the photon sphere.}
\label{table_b}
\end{center} 
\end{table}

\begin{figure}
\centering
\mbox{
(a)
\includegraphics[angle =0,trim=20mm 0mm 5mm 0mm, clip, scale=0.6]{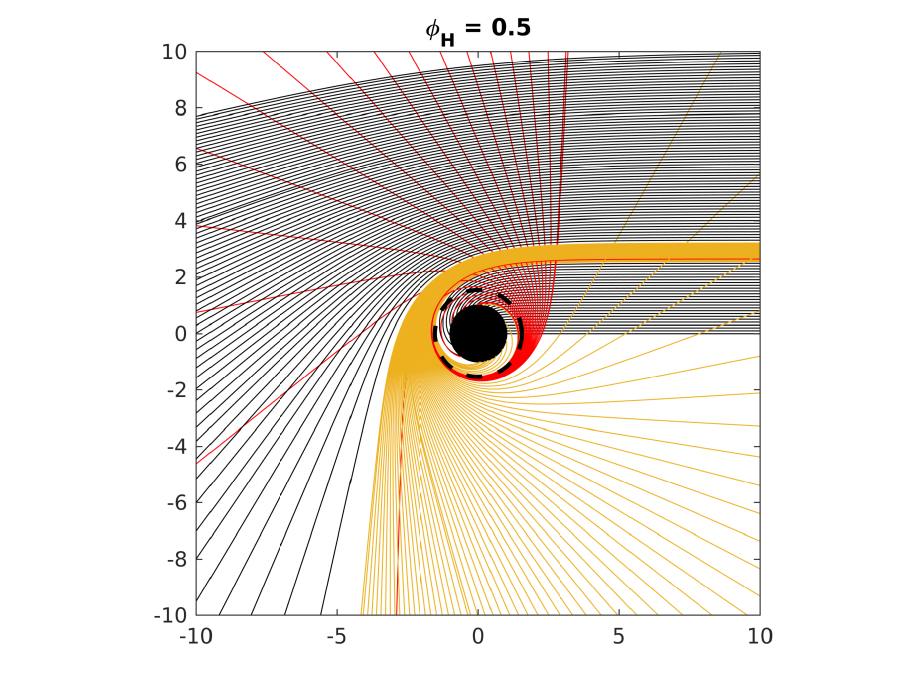}
(b)
\includegraphics[angle =0,scale=0.6]{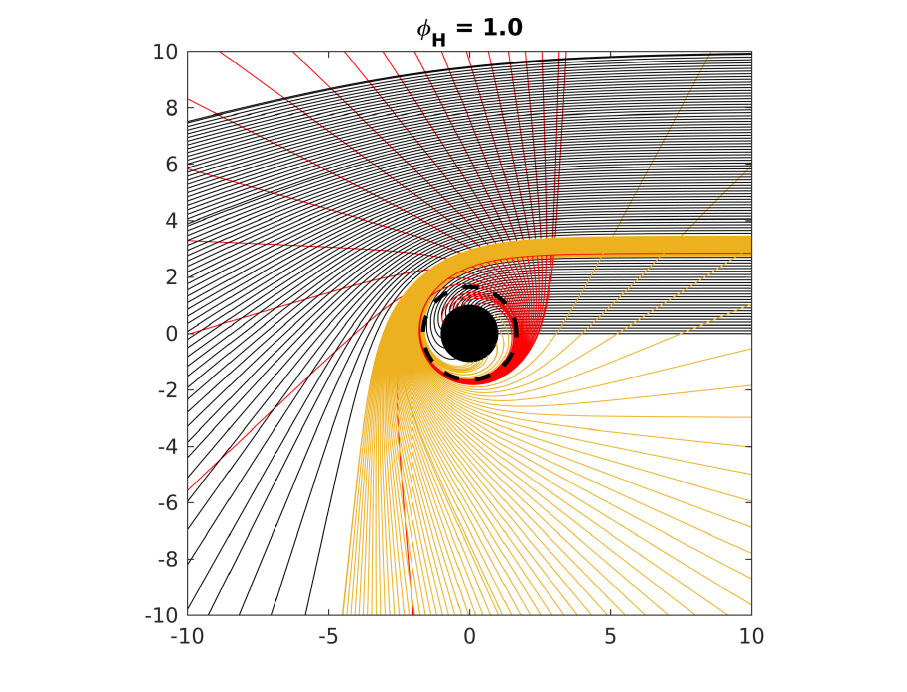}
}
\mbox{
(c)
\includegraphics[angle =0, trim=20mm 0mm 5mm 0mm, clip,scale=0.6]{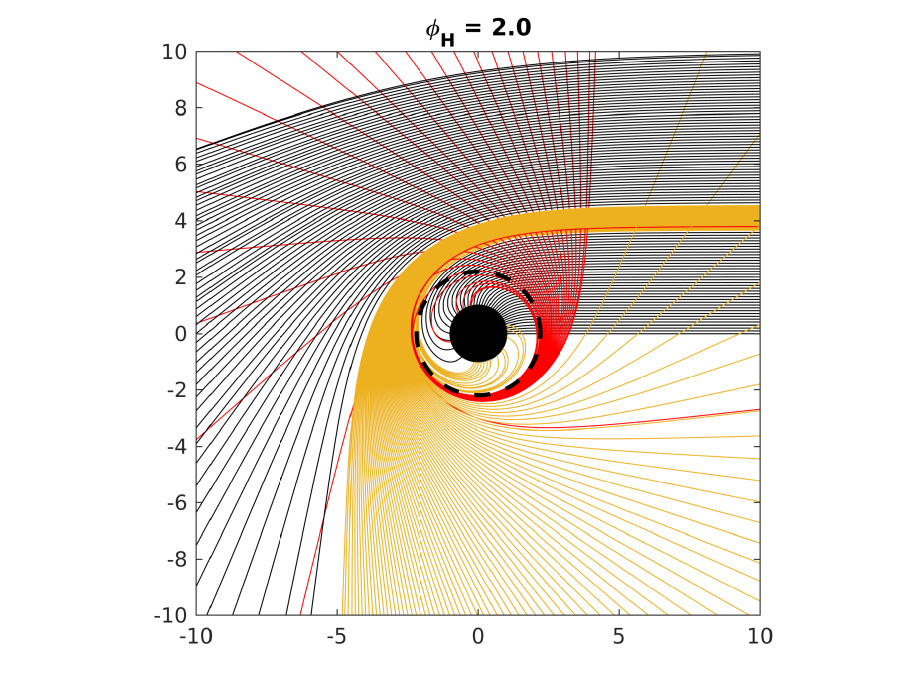}
(d)
\includegraphics[angle =0, trim=2.5mm 0mm 5mm 0mm, clip, scale=0.6]{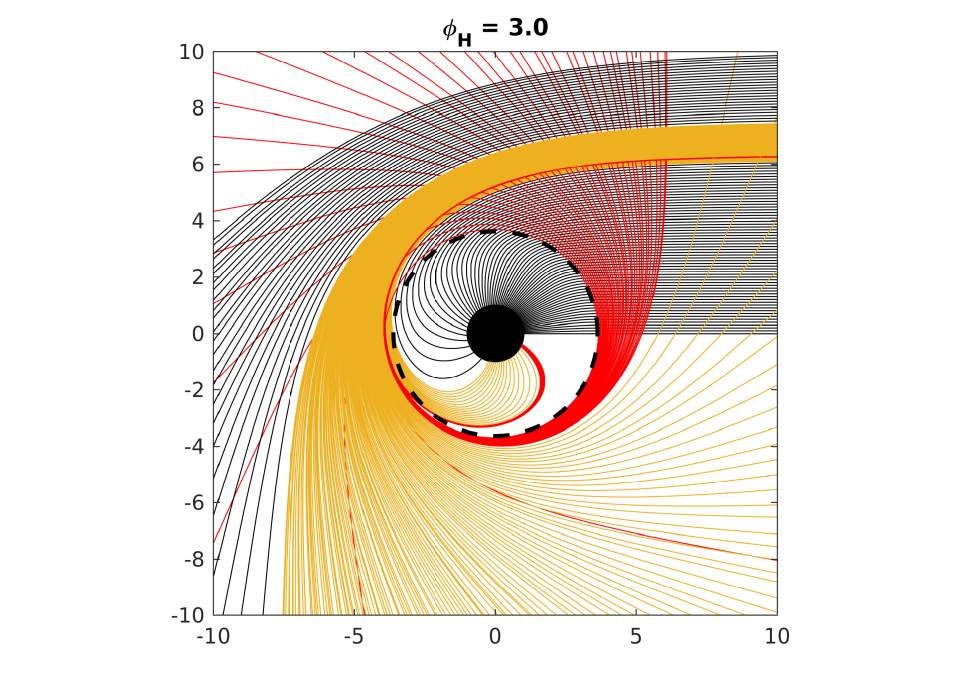}
}
\caption{
The trajectories of light rays are presented in the two-dimensional Cartesian coordinate $(x,y)$ for the HBHs with $r_H = 1$ and (a) $\phi_H=0.5$, (b) $\phi_H=1.0$, (c) $\phi_H=2.0$ and (d) $\phi_H=3.0$. A light ray is fixed by a value of impact parameter $b$, then the step size between the light rays for the direct (black), lensed (gold), and photon ring (red) emissions are fixed by 0.1, 0.01, and 0.001, respectively. The black holes are represented by the black disk. The photon sphere is represented by the dashed circle. The accretion disk lies on the vertical straight line with $(0,y)$.
}
\label{plot_trajectories}
\end{figure}

Next, the trajectories of light rays around the  HBHs with $r_H=1$ can also be visualized in the Cartesian coordinates $(x,y)$ by following the characterization of light rays $n=\varphi/(2\pi)$ in Fig.~\ref{plot_trajectories}. Nevertheless, some differences can be spotted for the light rays around these two types of black holes. Firstly, we can directly visualize the increase in the size of the photon sphere (dashed circle) in the vicinity of HBHs with the increment of $\phi_H$, where the size of the shadow for HBHs also increases with the increment of $\phi_H$. Secondly, the Table.~\ref{table_b} and Fig.~\ref{plot_orbits} have shown that $b^\pm_m$ and $b_c$ increase when $\phi_H$ increases, in particular we can clearly observe that not only the width between the lower ($b^-_2$) and upper ($b^+_2$) boundaries for the incoming lensed emission (gold) at the top right border of figures increases, but also their locations have been shifted upward. The trajectories of the photon ring (red) also exhibit this similar feature but are barely visible in the figures. Although these incoming light rays actually located further away from the HBHs, the falling and scattering of light rays around the HBHs can occur much earlier, since the deflection angle $\varphi$ is fixed by the characterization of $n$. In particular, for the case of $\phi_H=3.0$, the trajectories of the photon ring (red) and lensed (gold) cross the photon sphere (dashed circle), then fall directly into the HBH earlier when $b<b_c$, thus this leads to the existence of an empty gap inside the photon sphere (dashed circle) near the vicinity of the event horizon of the HBH, thus this portion of falling light rays into the HBH do not contribute to any observable trajectory in this region. 

\begin{figure}
\centering
\mbox{
(a)
\includegraphics[angle =0,scale=0.58]{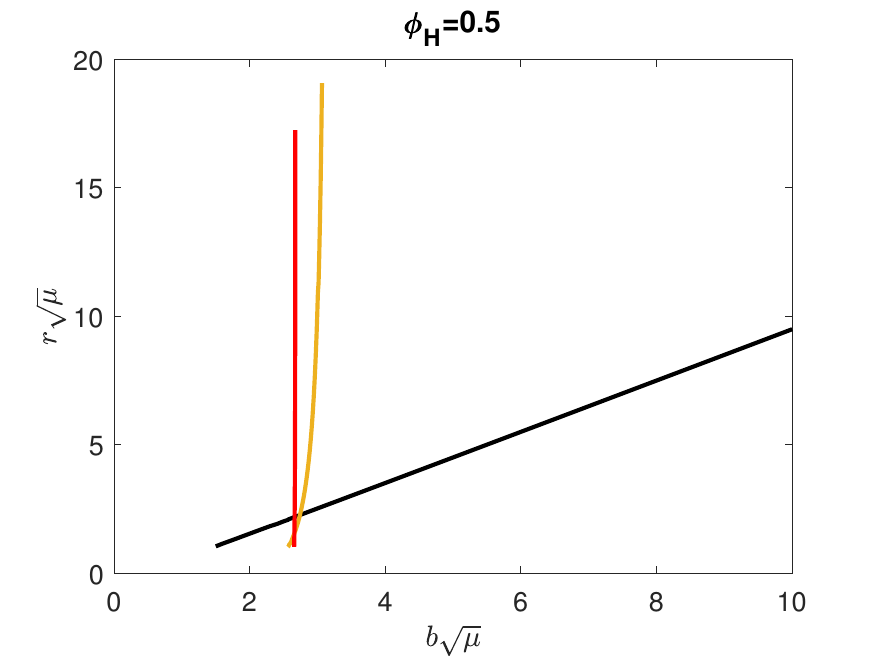}
(b)
\includegraphics[angle =0,scale=0.58]{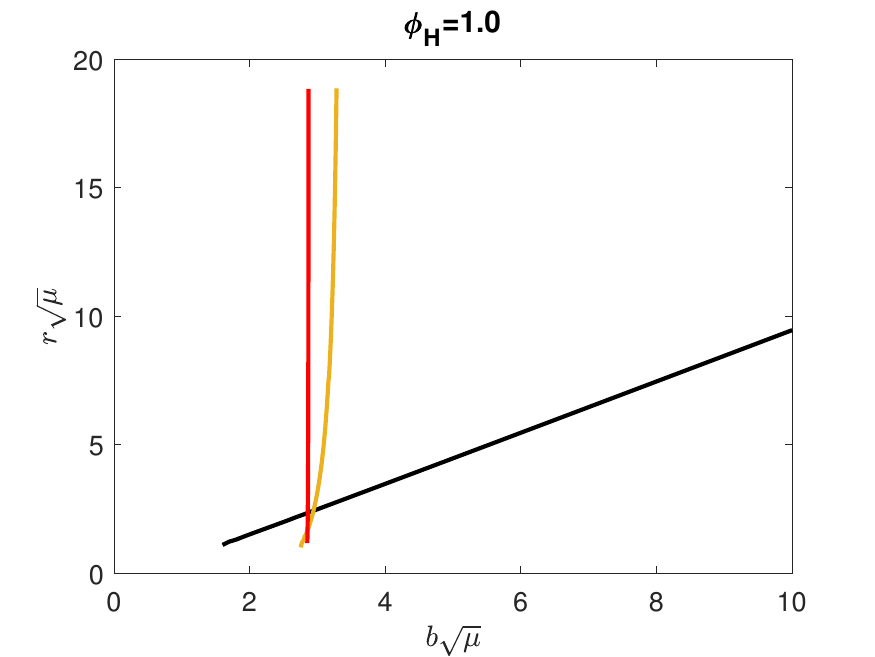}
}
\mbox{
(c)
\includegraphics[angle =0,scale=0.58]{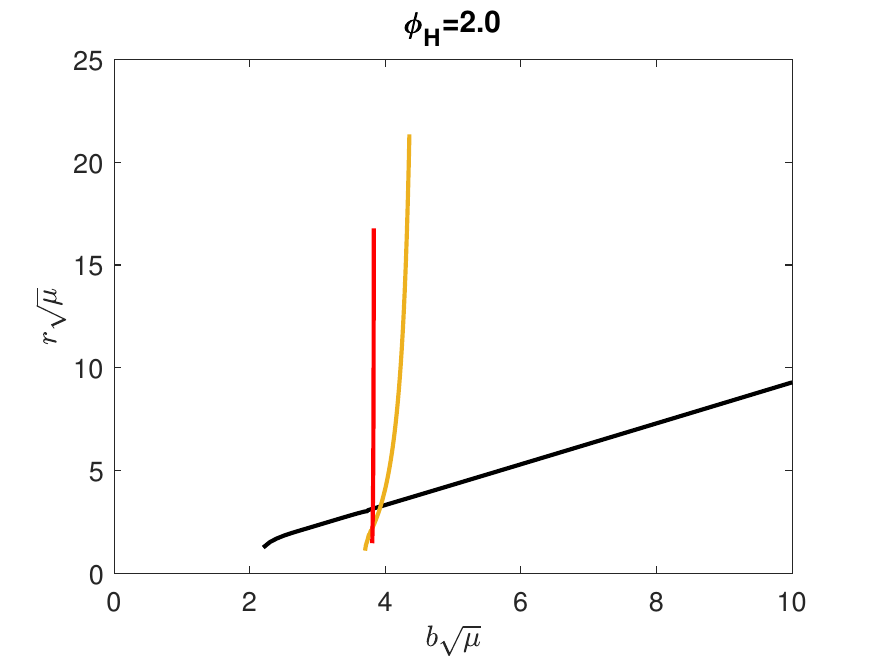}
(d)
\includegraphics[angle =0,scale=0.58]{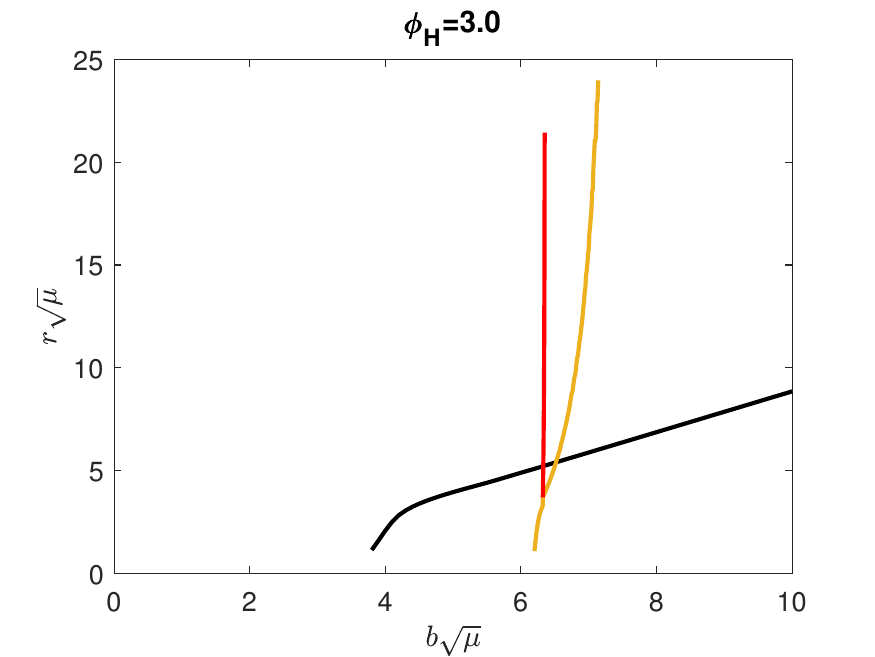}
}
\caption{The three transfer functions: $r_1(b)$ (black), $r_2(b)$ (gold), and $r_3(b)$ (red) as a function of scaled impact parameter $b\sqrt{\mu}$ for the HBHs with $r_H=1$, and (a) $\phi_H=0.5$, (b) $\phi_H=1.0$, (c) $\phi_H=2.0$ and (d) $\phi_H=3.0$.}
\label{plot_transfer_fun}
\end{figure}

\begin{figure}
\centering
\mbox{
(a)
\includegraphics[angle =0,scale=0.58]{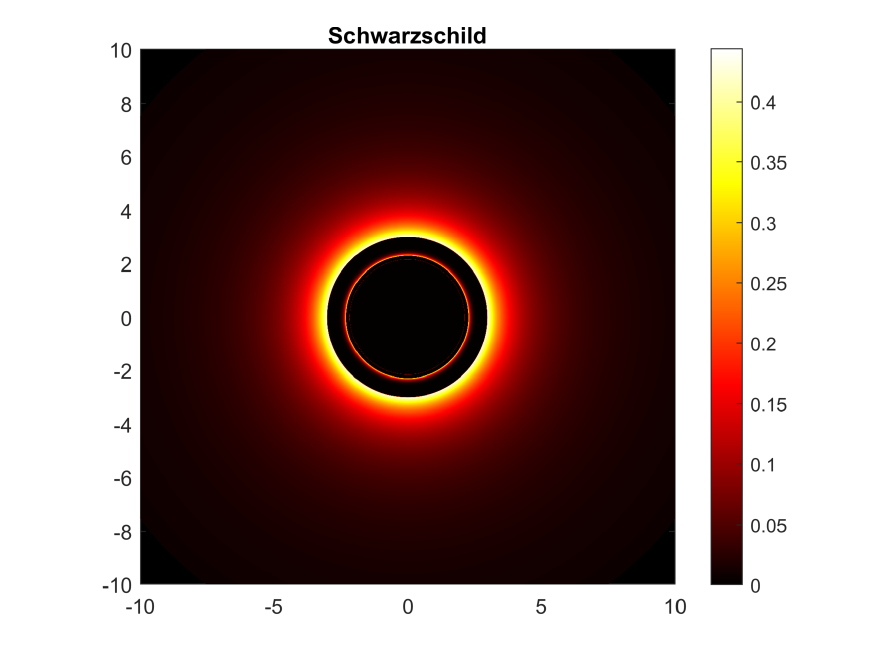}
(b)
\includegraphics[angle =0,scale=0.58]{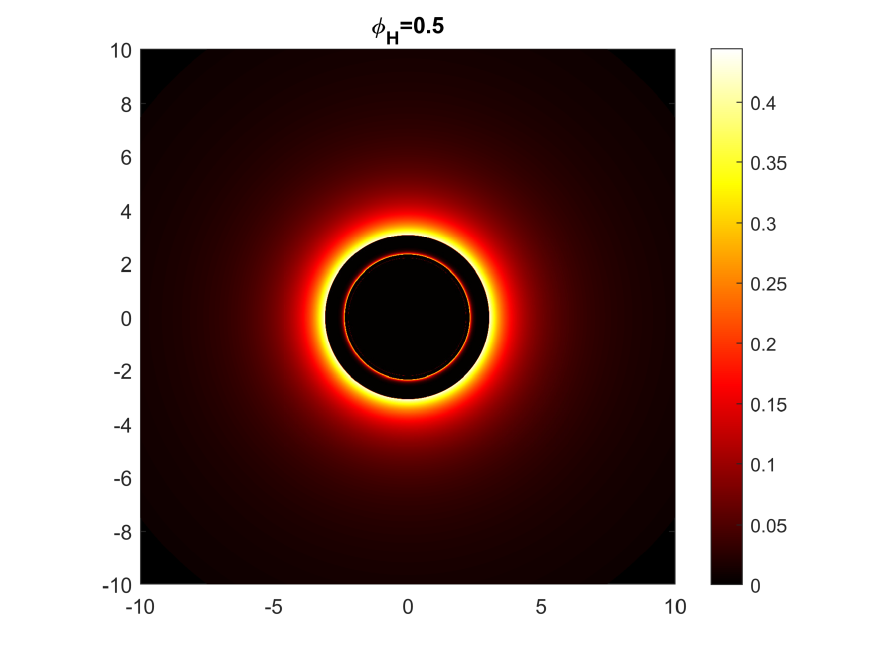}
}
\mbox{
(c)
\includegraphics[angle =0,scale=0.58]{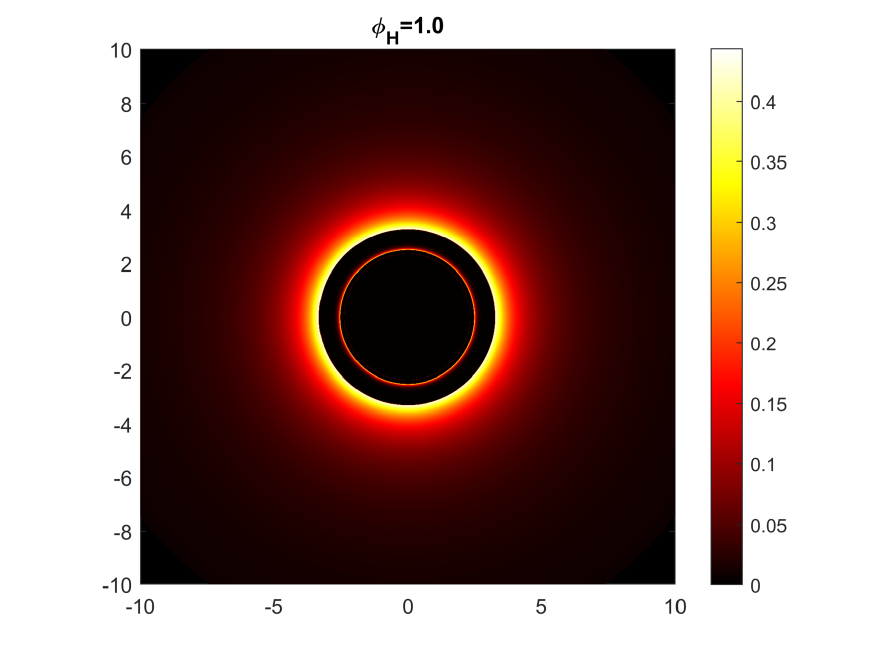}
(d)
\includegraphics[angle =0,scale=0.58]{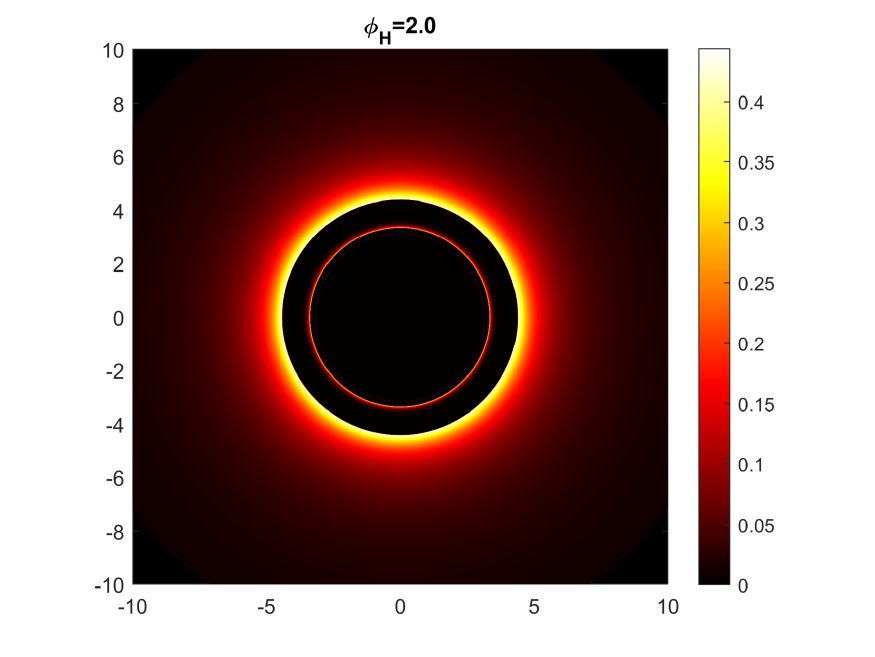}
}
\mbox{
(e)
\includegraphics[angle =0,scale=0.58]{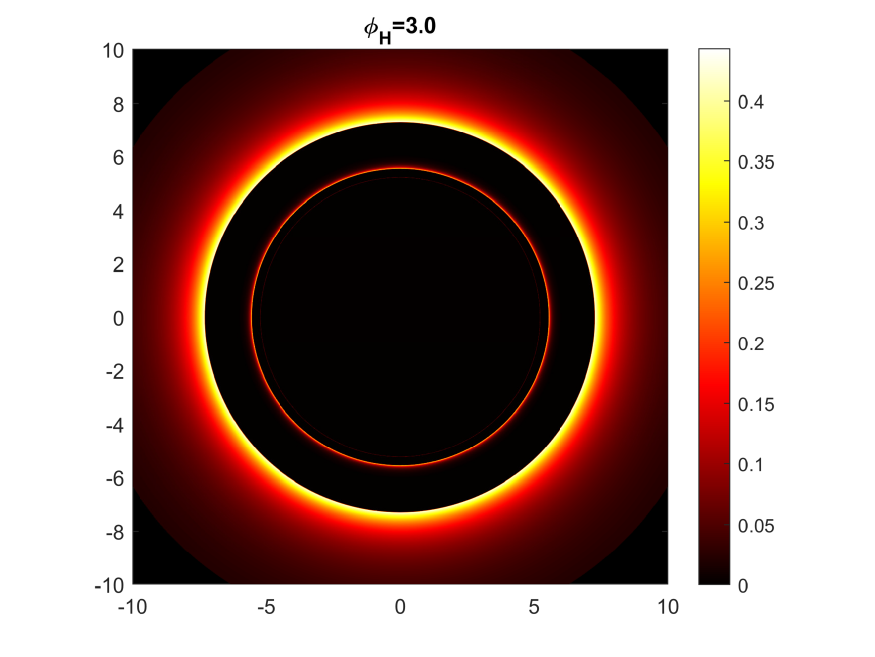}
}
\caption{
The optical appearance of Model I for (a) the Schwarzschild black hole; HBHs with $r_H=1$ and (b) $\phi_H=0.5$, (c) $\phi_H=1.0$, (d) $\phi_H=2.0$ and (e) $\phi_H=3.0$.}
\label{compare_Iobs1_optical}
\end{figure}

\begin{figure}
\centering
\mbox{
(a)
\includegraphics[angle =0,scale=0.58]{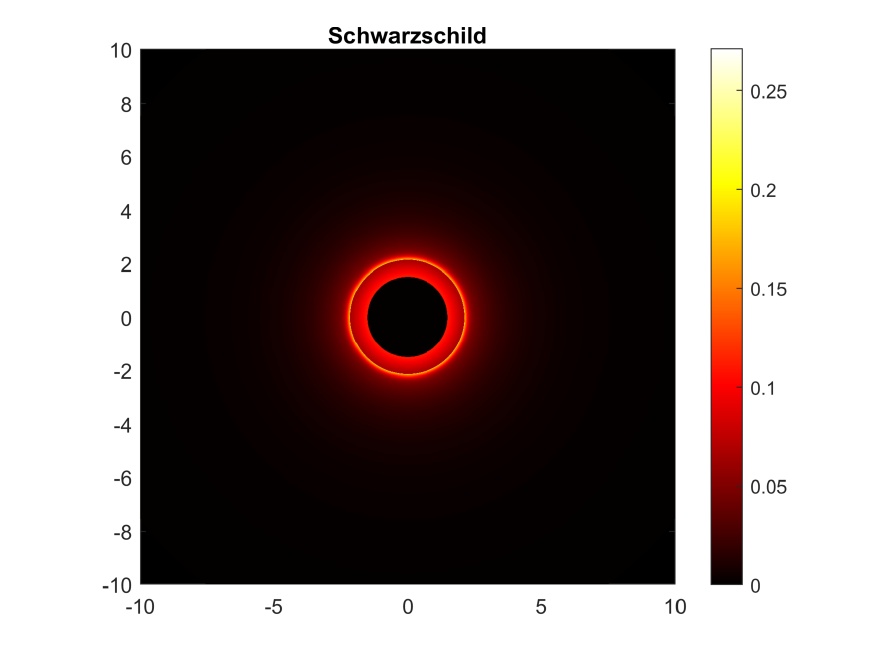}
(b)
\includegraphics[angle =0,scale=0.58]{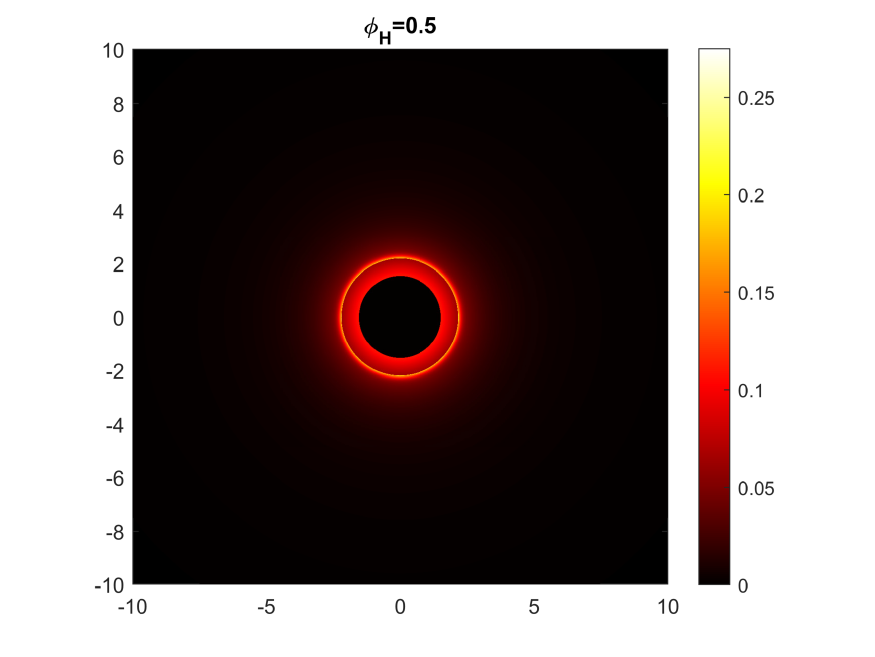}
}
\mbox{
(c)
\includegraphics[angle =0,scale=0.58]{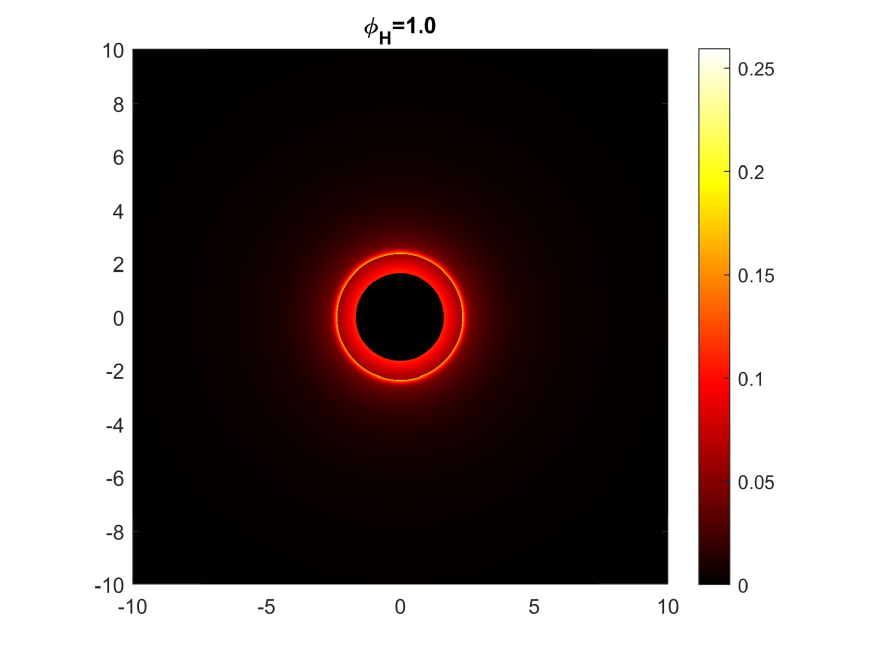}
(d)
\includegraphics[angle =0,scale=0.58]{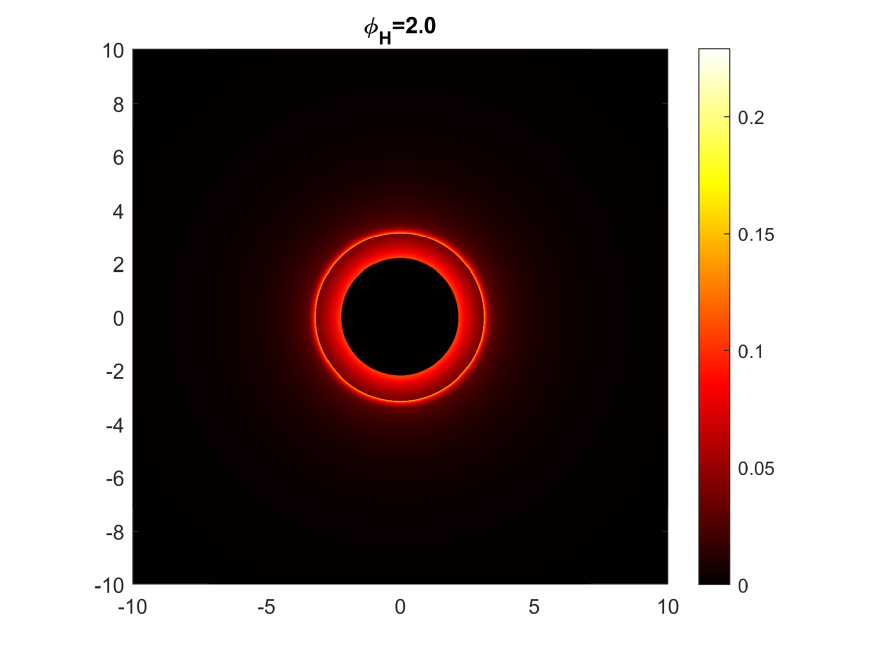}
}
\mbox{
(e)
\includegraphics[angle =0,scale=0.58]{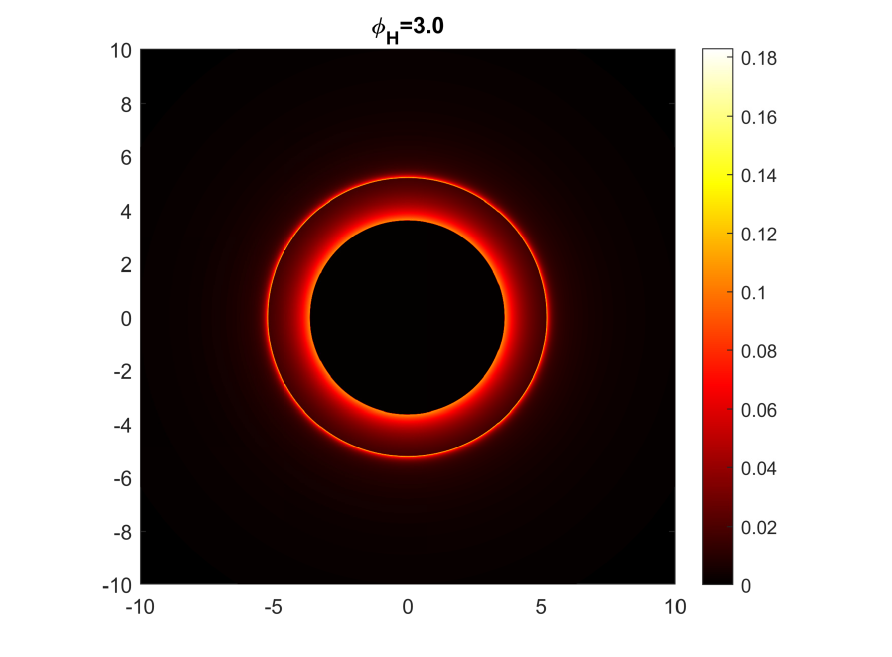}
}
\caption{
The optical appearance of Model II for (a) the Schwarzschild black hole; HBHs with $r_H=1$ and (b) $\phi_H=0.5$, (c) $\phi_H=1.0$, (d) $\phi_H=2.0$ and (e) $\phi_H=3.0$.
}
\label{compare_Iobs2_optical}
\end{figure}

\begin{figure}
\centering
\mbox{
(a)
\includegraphics[angle =0,scale=0.58]{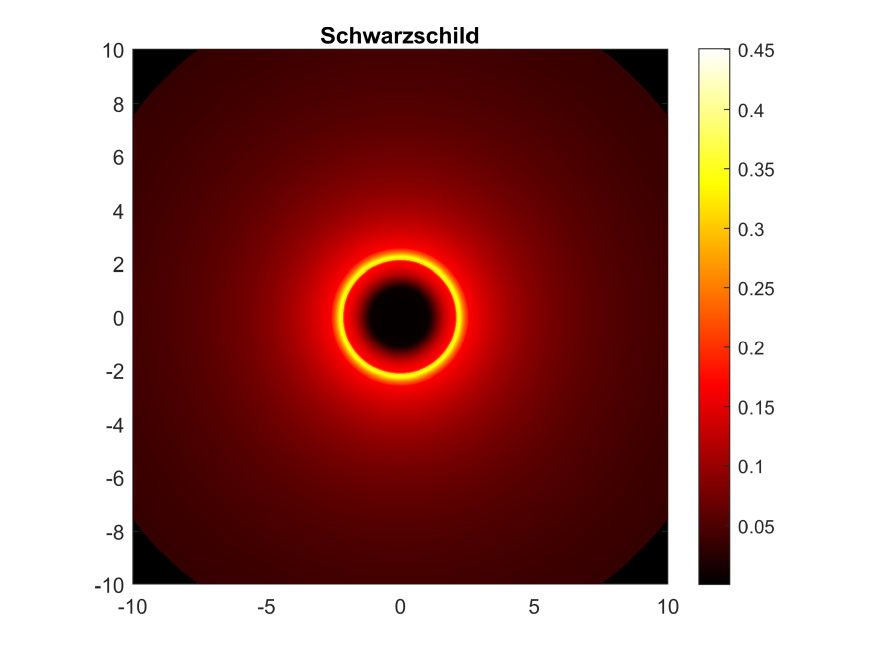}
(b)
\includegraphics[angle =0,scale=0.58]{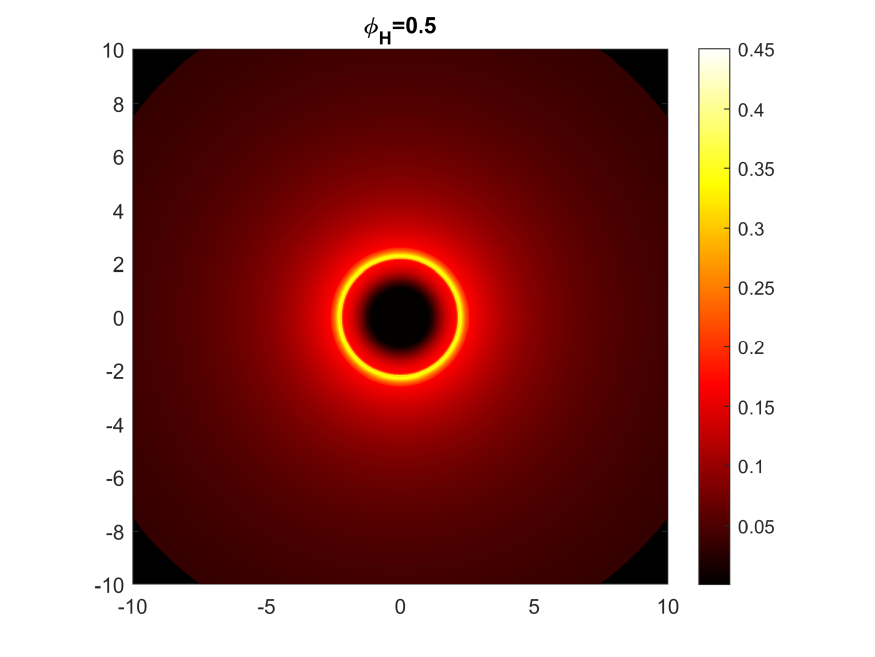}
}
\mbox{
(c)
\includegraphics[angle =0,scale=0.58]{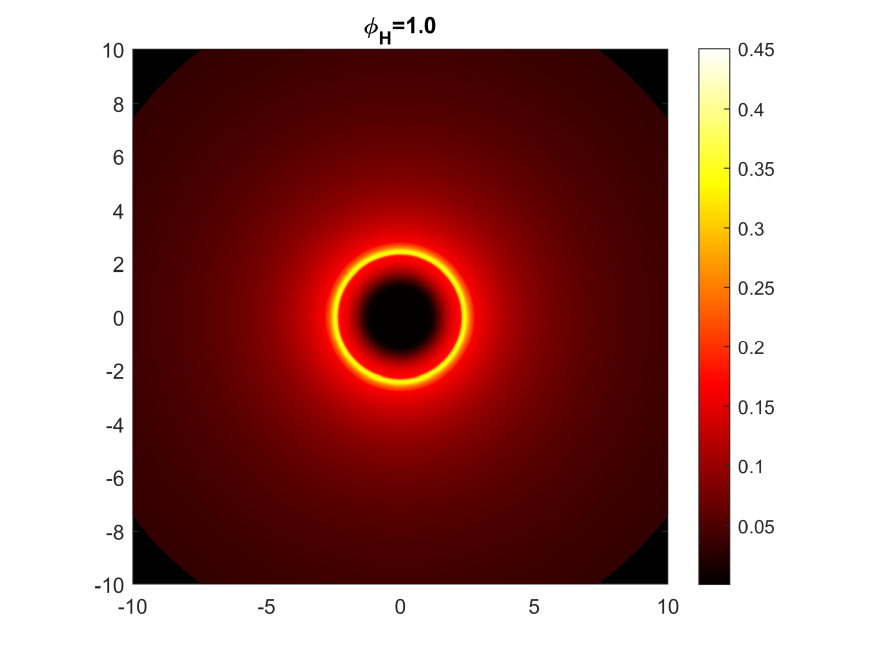}
(d)
\includegraphics[angle =0,scale=0.58]{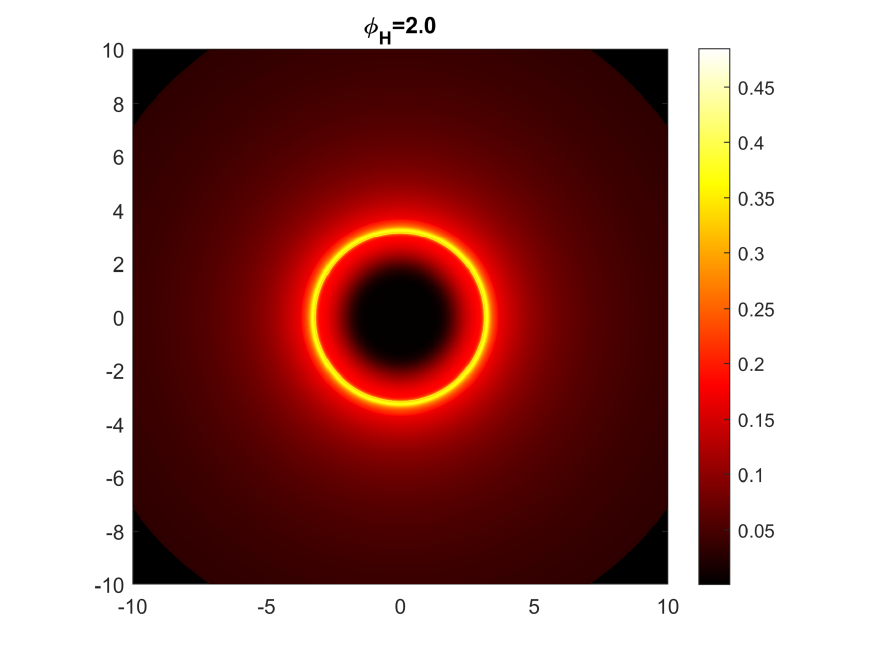}
}
\mbox{
(e)
\includegraphics[angle =0,scale=0.58]{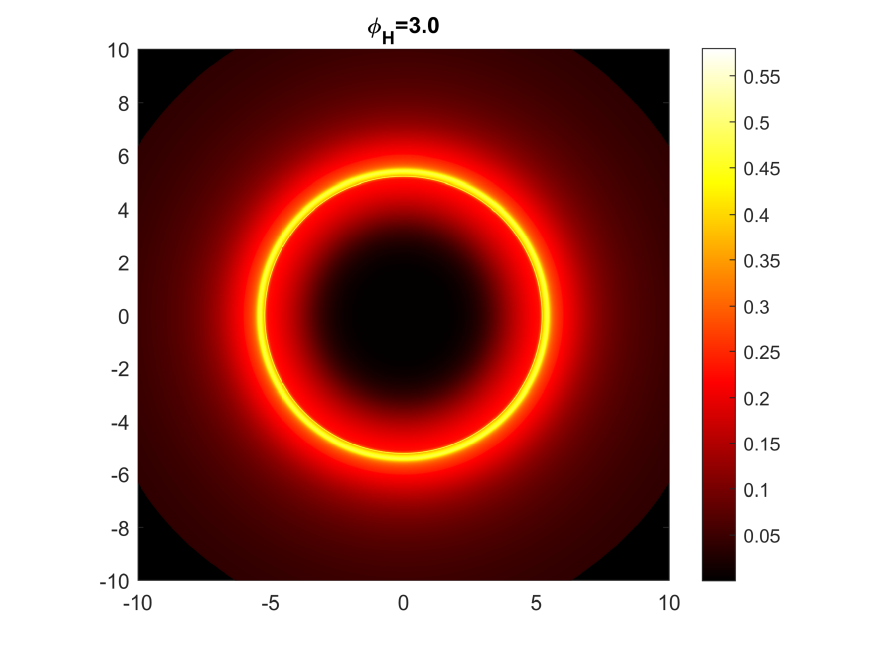}
}
\caption{
The optical appearance of Model III for (a) the Schwarzschild black hole; HBHs with $r_H=1$ and (b) $\phi_H=0.5$, (c) $\phi_H=1.0$, (d) $\phi_H=2.0$ and (e) $\phi_H=3.0$.
}
\label{compare_Iobs3_optical}
\end{figure}

Fig.~\ref{plot_transfer_fun} depicts the three transfer functions $r_1(b)$ (black), $r_2(b)$ (gold) and $r_3(b)$ (red) for direct, lensed and photon sphere emissions, respectively for the HBHs with $r_H=1$ and several $\phi_H$. When $\phi_H$ increases, all three transfer functions shift to the right, toward a higher value of $b$, significantly impacting the three trajectories of light rays by the following: Firstly, the right shift of $r_1(b)$ indicates that even minimally deflected light rays following the direct emission trajectory require a larger $b$ or larger distance to escape from the HBHs, since $b$ can also be defined as the distance of closest separation of the incoming light rays without deflection to the centre of a black hole.
Secondly, the positive horizontal shift of $r_2(b)$ causes the light rays in the lensed trajectory to revolve around the BH with greater paths due to the high $b$. Finally, the right shift of $r_3(b)$ indicates that the photon sphere itself is affected by the presence of $\phi_H$, with higher values of $\phi_H$ pushing the photon sphere outward. Note that for the higher values of $\phi_H$, such as 2.0 and 3.0, $r_m(b)$ increases very sharply when $b^{-}_m$ starts to increase, this indicates that the gravitational effects of the HBHs become more pronounced due to the expansion of the radius of the photon sphere as shown in Fig.~\ref{veff} (d) and Fig.~\ref{plot_trajectories}. This effect increases the slope $\frac{dr_m}{db}$ or the demagnification factor, resulting in highly demagnified images of the accretion disk for direct, lensed and photon ring emissions at lower $b$ and orbital radii. Notably, for $\phi_H=3.0$, the occurrence of an empty gap near the centre event horizon of the HBH in Fig.~\ref{plot_trajectories} (e) could potentially increase the slope or demagnification factor of the $r_m(b)$ at low $b$, and subsequently decreases the observed intensity. However, the effect is less obvious when translated to the optical appearance of the HBH with $\phi_H=3.0$ in Figs.~\ref{compare_Iobs1_optical}, \ref{compare_Iobs2_optical}, and  \ref{compare_Iobs3_optical} due to the small range of $b$ involved in the distorted slope as shown in Fig,~\ref{plot_transfer_fun}.

\begin{figure}
\centering
\mbox{
(a)
\includegraphics[angle =0,scale=0.58]{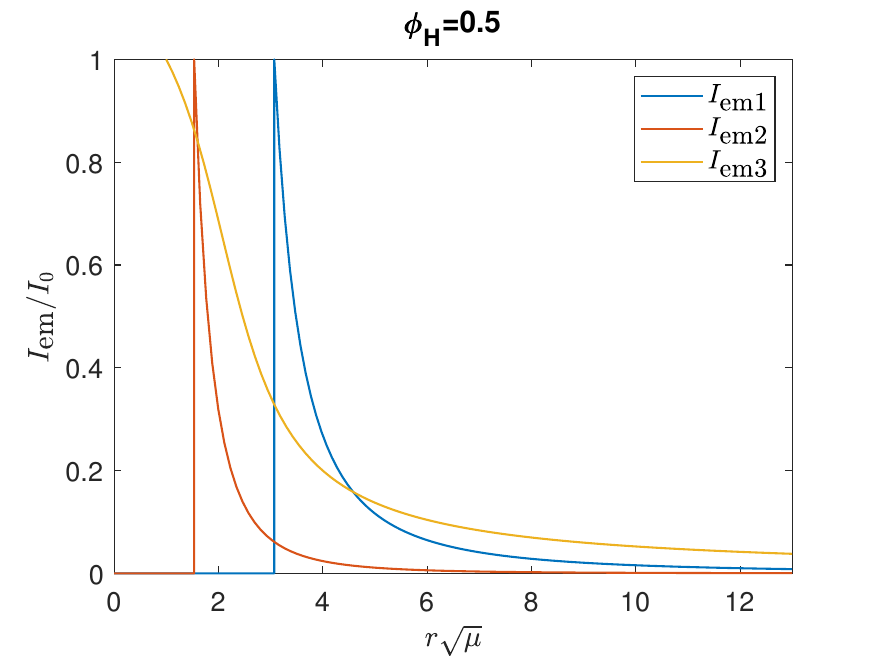}
(b)
\includegraphics[angle =0,scale=0.58]{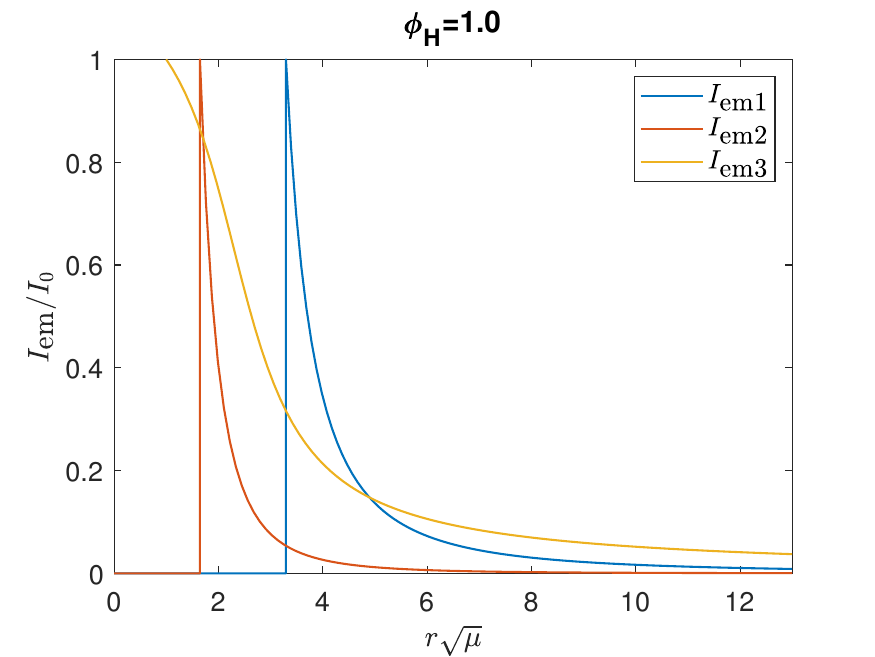}
}
\mbox{
(c)
\includegraphics[angle =0,scale=0.58]{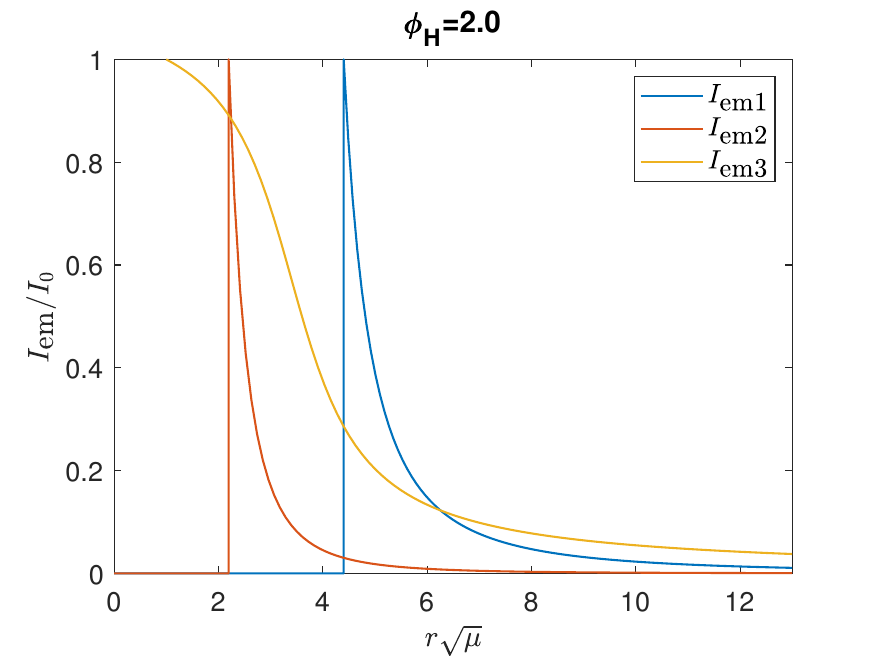}
(d)
\includegraphics[angle =0,scale=0.58]{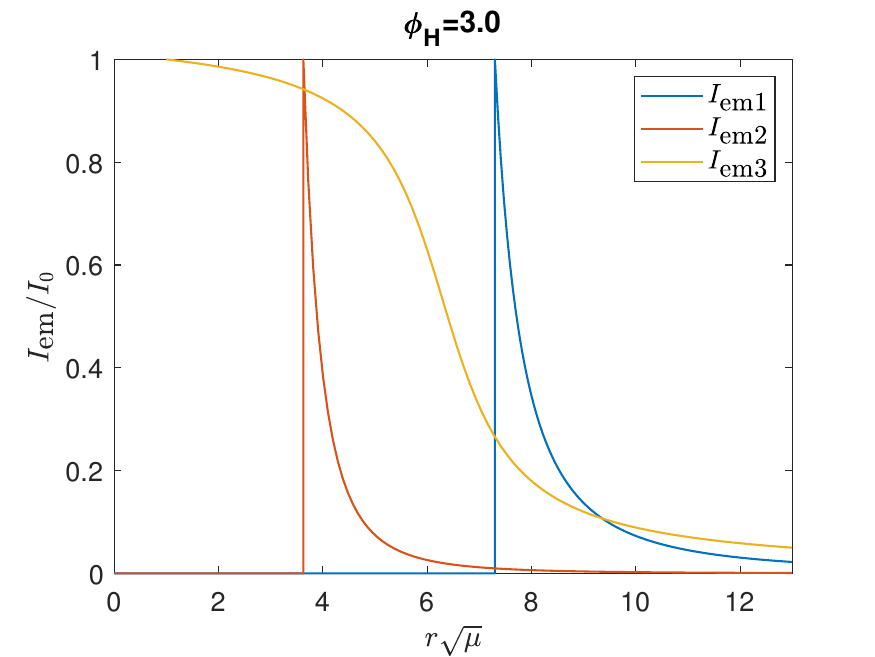}
}
\caption{Three models of emitted intensity profiles: $I_\text{em1}(r)/I_0$, $I_\text{em2}(r)/I_0$, $I_\text{em3}(r)/I_0$ as a function of scaled $r\sqrt{\mu}$ for the HBHs with $r_H=1$ and (a) $\phi_H=0.5$, (b) $\phi_H=1.0$, (c) $\phi_H=2.0$ and (d) $\phi_H=3.0$.}
\label{phi_H_Iem_combine}
\end{figure}

\begin{figure}
\centering
\mbox{
(a)
\includegraphics[angle =0,scale=0.58]{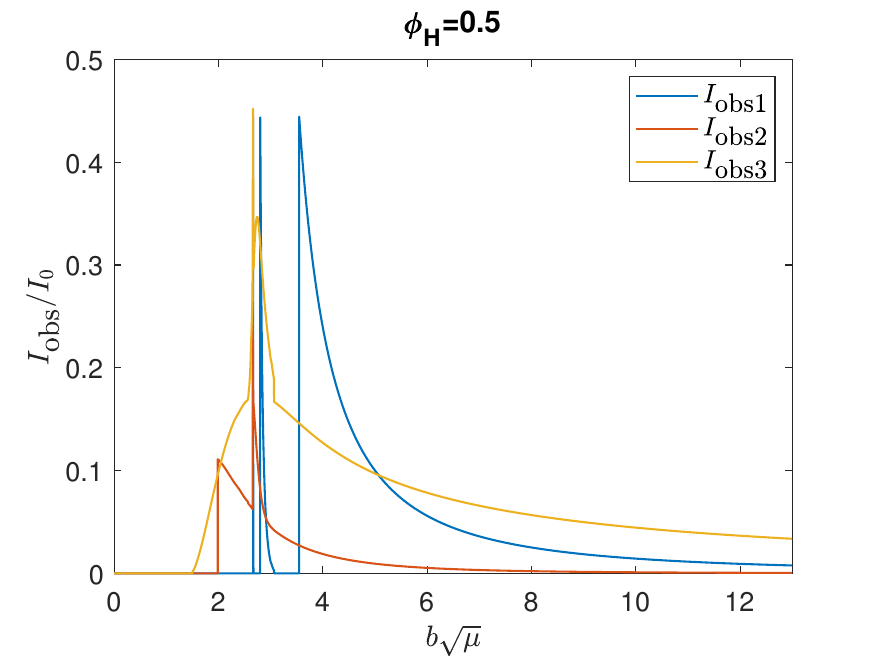}
(b)
\includegraphics[angle =0,scale=0.58]{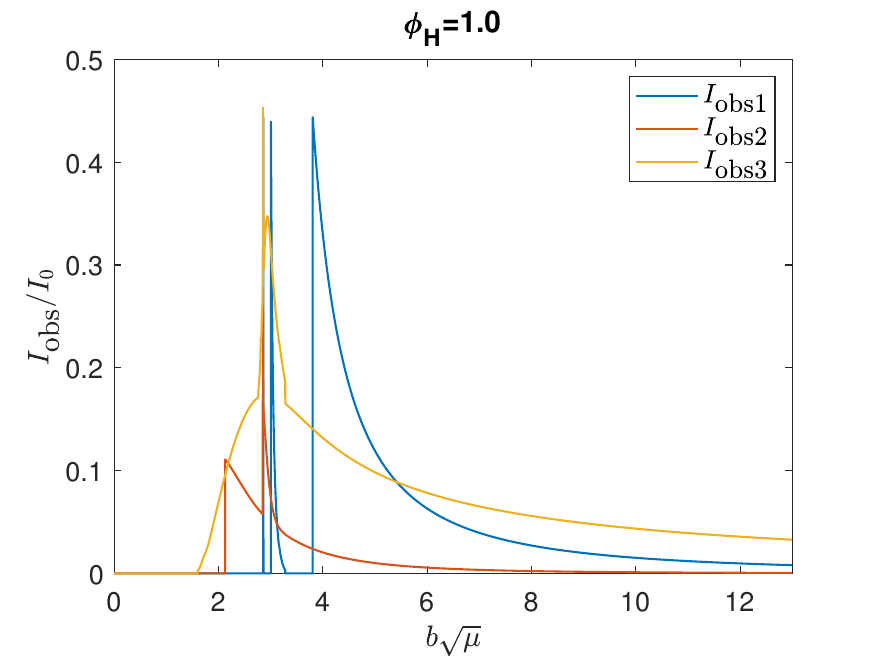}
}
\mbox{
(c)
\includegraphics[angle =0,scale=0.58]{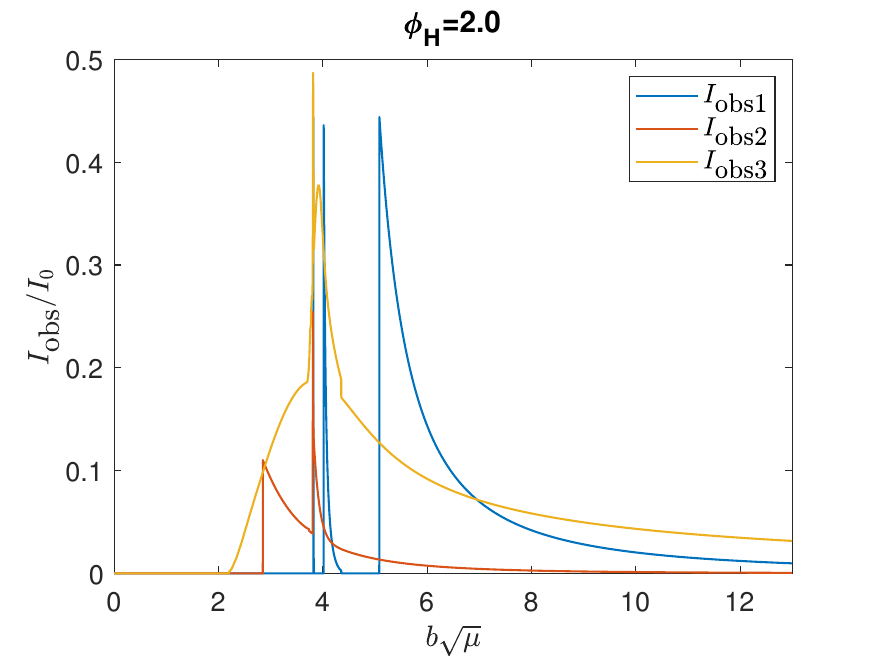}
(d)
\includegraphics[angle =0,scale=0.58]{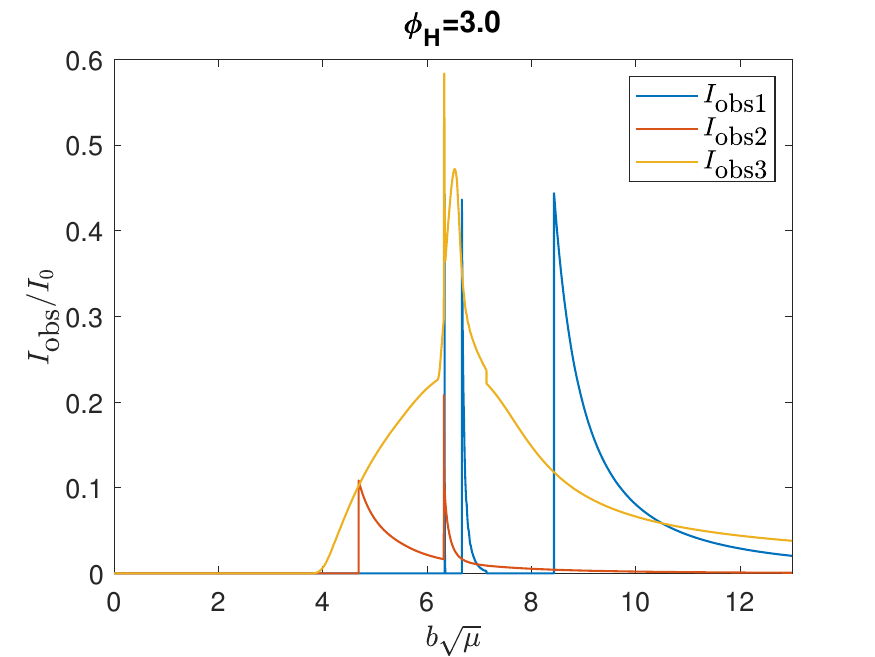}
}
\caption{
Three models of observed intensity profiles: $I_\text{obs1}(r)/I_0$, $I_\text{obs2}(r)/I_0$, $I_\text{obs3}(r)/I_0$ as a function of scaled $r\sqrt{\mu}$ for the HBHs with $r_H=1$ and (a) $\phi_H=0.5$, (b) $\phi_H=1.0$, (c) $\phi_H=2.0$ and (d) $\phi_H=3.0$.}
\label{phi_H_Iobs_combine}
\end{figure}

Fig.~\ref{phi_H_Iem_combine} shows three models of the emitted intensity of light rays $I_\text{em}(r)/I_0$ from the accretion disk around the HBHs with $r_H=1$. We observe $I_\text{em1}(r)/I_0$, $I_\text{em2}(r)/I_0$ and $I_\text{em3}(r)/I_0$ for the Schwarzschild black hole and HBHs behave quite similar.  Nevertheless, since Fig.~\ref{veff} shows that $r_\text{ISCO}$ and $r_p$ increase monotonically as $\phi_H$ increases, thus the emissions of $I_\text{em1}(r)/I_0$ and $I_\text{em2}(r)/I_0$ 
begin at a larger distance of $r$ for the HBHs. Although we fix the size of the horizon for the Schwarzschild black hole and HBHs as $r_H=1$, $I_\text{em3}(r)/I_0$ decreases more slowly to zero when $\phi_H$ increases.

Fig.~\ref{phi_H_Iobs_combine} exhibits the total luminosity of light rays $I_\text{obs}(r)/I_0$ for the HBHs with $r_H=1$ received by an observer at infinity. We focus our discussion on Model I, which is $I_\text{obs1}(r)/I_0$. When $\phi_H>0$, $I_\text{obs1}(r)/I_0$ is still dominated by the direct emission, which corresponds to the outermost spike of $I_\text{obs1}(r)/I_0$. However, the separation between the spikes of $I_\text{obs1}(r)/I_0$ also increases. In  Fig.~\ref{compare_Iobs1_optical}, the optical appearance of the Schwarzschild black hole and HBHs behave qualitatively similar, where they possess the brightest ring at the outermost due to the direct emission, a dimmer lensed ring due to the lensed emission and a faint ring in the innermost region due to the photon ring emission. Note that the photon ring can be seen by zooming in on the figure. Since all the transfer functions (Fig. \ref{plot_transfer_fun}), $I_\text{em1}(r)/I_0$ (Fig.~\ref{phi_H_Iem_combine}), and $I_\text{obs1}(r)/I_0$ (Fig.~\ref{phi_H_Iobs_combine}) exhibit a rightward shift as $\phi_H$ increases from zero, these lead to the expansion of the HBH shadows and accretion disks as shown in Fig. \ref{compare_Iobs1_optical}.

Although the emission profile of Model II for the Schwarzschild black hole and HBHs is qualitatively the same, the direct emission for the HBHs occurs at larger $r$ since $r_p$ increases when $\phi_H$ increases. As a result, the inner edge of the accretion disk in Fig. \ref{compare_Iobs2_optical} lies on the critical curve or shadow. The redshift effects can reduce the observed flux. The optical images of the Schwarzschild black hole and HBHs show the lensing ring more clearly, but the photon ring is barely visible. Similar to Model I, the shadows and accretion disks of Model II expand as $\phi_H$ increases. The expansion of the area under the graph $I_\text{obs2}(r)/I_0$ (Fig.~\ref{phi_H_Iobs_combine}), further justifies the increase in accretion disk size.

In Model III, the enlargement of the shadows and accretion disks is also noticeable as $\phi_H$ increases. The inner edge of the accretion disk extends down to the event horizon, resulting in a broader region of luminosity. The lensing ring and photon ring again overlap on the direct image, with the lensed ring becoming more pronounced. However, direct emission still dominates most of the overall brightness. The optical images of the Schwarzschild black hole and HBH reveal a wider area of brightness, clearly showing the lensed ring and a fainter photon ring near the inner edge. However, the photon ring's contribution to the overall brightness remains minimal.

\begin{figure}
\centering
\mbox{
(a)
\includegraphics[angle =0,scale=0.58]{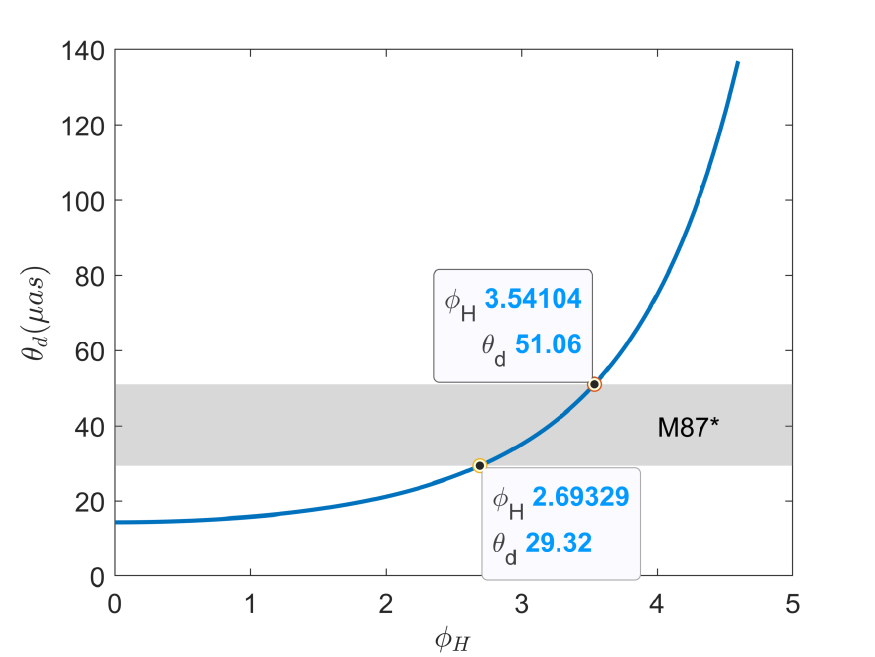}
(b)
\includegraphics[angle =0,scale=0.58]{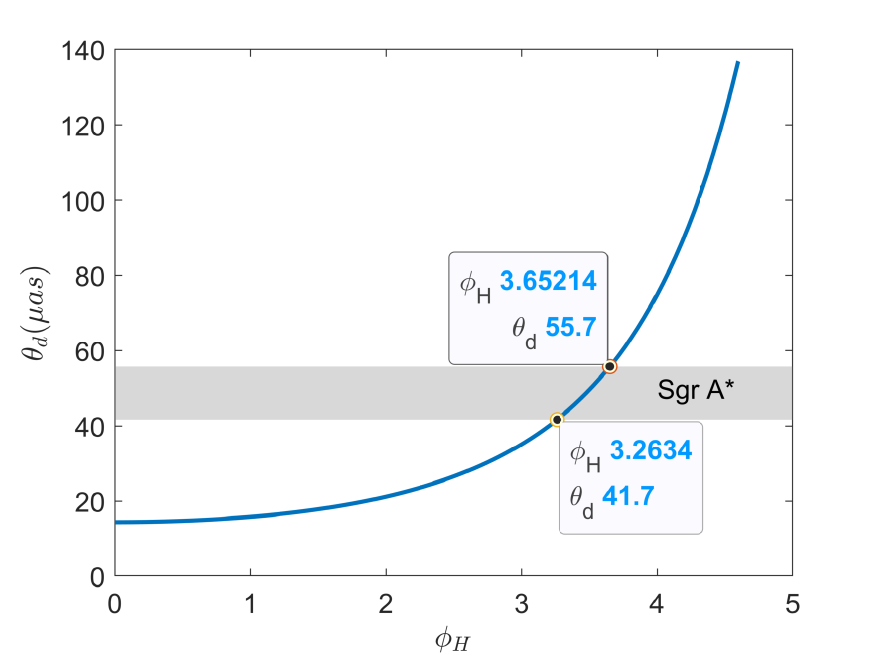}
}
\caption{The constraints imposed on $\phi_H$ by the observational data on the angular diameter $\theta_d$ of supermassive black holes in (a) M87 and (b) Sgr A$^{*}$. The two dots represent the upper and lower bounds.}
\label{theta_d_constraints}
\end{figure}

Finally, we constraint the parameter $\Lambda$ in $V(\phi)$ based on the observational data from the EHT on the angular diameter of shadow $\theta_d$ of the supermassive black holes M87 and  Sgr A$^{*}$, which is given by \cite{Kumar:2020owy, Wang:2024lte}:
\begin{equation}
    \theta_d = 2 \frac{b_c/\sqrt{\mu}}{d_L}\,,
\end{equation}
where $d_L$ represents the distance between the black hole and the distant observer. $\theta_d$ for the supermassive black hole in M87* with mass $M = 6.5 \times 10 M_\odot$ and $d_L = 16.8 \mbox{Mpc}$ is estimated to lie between $29.32 \mu as$ and $51.06 \mu as$ \cite{EventHorizonTelescope:2021dqv}. In contrast, $\theta_d$ for Sgr A$^{*}$ with a mass of $M=4.0 \times 10^6 M_\odot$ and a distance of $d_L=8.15 \mbox{kpc}$ is $\theta_d = 48.7 \pm 7.0 \mu as$ \cite{EventHorizonTelescope:2022wkp, EventHorizonTelescope:2022xqj}. Note that the mass of a scalar field $m_s$ is given by $m_s=\sqrt{2\mu}$ which is assumed to be a ultralight boson with mass $4 \times 10^{-20}$ eV \cite{Cunha:2019ikd}. Thus, these data also allow us to infer constraints on $\phi_H$. As  depicted in Fig.~\ref{theta_d_constraints}, the range of $\Lambda$ is $0.472984 \leq \Lambda \leq 0.523082$ corresponds to $3.65214 \leq \phi_H \leq 3.2634$ for Sgr A$^{*}$ and $0.485277 \leq \Lambda \leq 0.649827$ corresponds to $3.54104 \leq \phi_H \leq 2.69329$ for M87.

\section{Conclusion}\label{sec:con}

In conclusion, we study the optical appearance of a hairy black hole (HBH), which can be constructed via a minimal coupling of Einstein gravity with a non-positive definite scalar potential $V(\phi)=-\Lambda \phi^4 + \mu \phi^2$ in the Einstein-Klein-Gordon theory. When the scalar field $\phi_H$ at the event horizon $r_H$ increases, the HBH with the corresponding $\Lambda$ can be bifurcated from the Schwarzschild black hole by exhibiting some new phenomena. Besides, we fix the size of horizon $r_H$ for the Schwarzschild black hole and HBH as $r_H=1$. 

Hence, we adopt the ray-tracing method to study the optical appearance of the HBHs under the influence of $\phi_H$. In this framework, we assume the HBH is surrounded by an optically and geometrically thin accretion disk on the equatorial plane, where the trajectories of light rays can be fully described by the radial geodesics equation. Hence, we derive the effective potential for massive test particles and light rays. We find that the location of the innermost of stable circular orbit $r_\text{ISCO}$ and photon sphere $r_p$ increase monotonically as $\phi_H$ increases.

Analogous to the Schwarzschild black hole, the trajectories of light rays around the HBH can be characterized into three emissions with some specific range of the impact parameter $b$: direct, lensed and photon ring. However, the range of $b$ for a specific emission increases when $\phi_H$ increases, exhibiting a significant deviation from the Schwarzschild black hole. Therefore, the transfer function of three emissions also shifted to higher values of $b$ for the HBHs, then deviated significantly from the Schwarzschild black hole, in particular for large $\phi_H$, these functions increase sharply when $b$ increases from its lower boundary.

We have adopted three models of emitted and observed intensity with three different starting points of emission which are $r_\text{ISCO}$, $r_p$ and $r_H$ to describe the optical appearances of HBH. When $\phi_H$ increases, we find that the profiles of emitted and observed intensity for the HBH and Schwarzschild black hole behave qualitatively the same but the starting points of emission have been increased. The size of the bright rings and shadow increases with increasing $\phi_H$ compared to the Schwarzschild black hole. However, the brightness of the rings still remains almost affected, this implies that the HBH can potentially become a black hole mimicker if we vary $r_H$, since we have fixed $r_H=1$ in the calculation. Furthermore, we have constrained the parameter $\Lambda$ in $V(\phi)$ based on the observations from the Sgr A$^{*}$ and M87 as $0.472984 \leq \Lambda \leq 0.523082$ and $0.485277 \leq \Lambda \leq 0.649827$, respectively.

Lastly, since the accretion disks surrounding the supermassive black holes are generally optically thin but geometrically thick, thus in the future, it would be interesting to extend our current work by investigating the accretion process for the HBH surrounded by such accretion disk, namely the ``Polish doughnuts" \cite{Pugliese:2012ub,Abramowicz:2011xu}.

\section*{Acknowledgement}
 XYC is supported by the starting grant of Jiangsu University of Science and Technology (JUST). We acknowledge having useful discussions with Minyong Guo, Hyat Huang, Zhen Li and Zi-liang Wang.


\bibliography{mybiblio}

\end{document}